\documentclass[]{article}% insert '[draft]' option to show overfull boxes
\usepackage{graphicx}
\usepackage{epsfig}
\usepackage{epstopdf}
\usepackage{authblk}
\usepackage{subfigure}% subcaptions for subfigures
\usepackage{wrapfig}%   wrap figures/tables in text (i.e., Di Vinci style)
\usepackage[T1]{fontenc}
\usepackage[latin9]{inputenc}
\usepackage{units}
\usepackage{amsmath}
\usepackage{amssymb}
\usepackage{esint}
\usepackage{float}
\usepackage{longtable,tabularx}
\usepackage{color}
\title{{\normalsize under consideration for publication in Philosophical Transactions A} \vspace {7mm} \\
 Modeling and prediction of the peak radiated sound in sub-sonic axisymmetric air jets using acoustic analogy based asymptotic analysis}

\author[1]{Mohammed Z. Afsar}
\author[2]{Adrian Sescu}
\author[3]{Stewart. J. Leib}

\affil[1]{Department of Mechanical and Aerospace Engineering, University of Strathclyde, Glasgow, UK}
\affil[2]{Department of Aerospace Engineering, Mississippi State University, Starkville, MS 39762}
\affil[3]{Ohio Aerospace Institute, Cleveland, OH 44142, USA.}

\begin{document}

\date{}

\maketitle

\begin{abstract}
This paper uses asymptotic analysis within the generalized acoustic analogy formulation (Goldstein. \emph{JFM} 488, pp. 315-333, 2003) to develop a noise prediction model for the peak sound of 
%high Reynolds number 
axisymmetric round jets at subsonic acoustic Mach numbers ($Ma$).
%how self-consistent asymptotic analysis within
The analogy shows that the exact formula for the acoustic pressure is given by a convolution product of a propagator tensor (determined by the vector Green's function of the adjoint linearized Euler equations
for a given jet mean flow) and a generalized source term representing the jet turbulence field.
%is a stationary random function of time 
%whose auto-covariance we assume is known, the  acoustic propagation is considered separately from the source modeling. 
%Since the analogy shows that the acoustic pressure is given by an exact convolution product between a propagator tensor (that depends on the jet mean flow via the vector Green's function of the adjoint linearized Euler equations) and a generalized source term that is a stationary random function of time whose auto-covariance we assume is known, the  acoustic propagation is considered separately from the source modeling. 
%
%In this paper we use asymptotic analysis within the generalized acoustic analogy formulation (Goldstein. J. Fluid Mech. 488, pp. 315-333, 2003) to develop a robust noise prediction model for  axisymmetric round jets at $O(1)$ subsonic acoustic Mach numbers.

Using a low frequency/small spread rate asymptotic expansion of the propagator, 
%scaling (Goldstein {\it et al}. \emph{JFM} 695, p.199, 2012) to determine the propagator at frequencies of the order of the small jet spread rate,  
mean flow non-parallelism enters the lowest order Green's function solution via the streamwise component of the mean flow advection vector in a hyperbolic partial differential equation (PDE). 
%Because this result is everywhere different from that obtained by a locally parallel flow, comparison of predictions with the former reveal  (among other things) that the non-parallelism amplifies the low frequency sound.
%
We then address the predictive capability of the solution to this PDE when used in the analogy through first-of-its-kind numerical calculations when an experimentally-verified model of the turbulence source structure is used together with Reynolds-averaged Navier Stokes solutions for the jet mean flow.
%
%Our noise predictions show a reasonable level of accuracy at high $Ma$, for example up to a Strouhal number of about $0.6$ for polar observation angles in the peak noise direction at $Ma=0.9$. 
%
Our noise predictions show a reasonable level of accuracy in the peak noise direction at $Ma=0.9$, for Strouhal number up to about $0.6$, and at $Ma=0.5$ using modified source coefficients. Possible reasons for this are discussed. 
%At lower acoustic Mach numbers the prediction require greater empirical tuning of the appropriate turbulence correlation indicating that 
%when an experimentally-verified model of the turbulence source structure is used together with Reynolds-averaged Navier Stokes solutions for the jet mean flow.
%
Moreover, the prediction range can be extended beyond unity Strouhal number by using an approximate composite asymptotic formula for the vector Green's function that reduces to the locally parallel flow limit at high frequencies.

\end{abstract}
%%%%%%%%%%%%%%%%%%%%%%%%%%%

%%%%%%%%%% Insert the texts which can accomdate on firstpage in the tag "fmtext" %%%%%

%\begin{fmtext}
%\newpage

% Biot savart law. FFwH. Saffman here.
%

%\end{fmtext}

%%%%%%%%%%%%%%% End of first page %%%%%%%%%%%%%%%%%%%%%

%\maketitle
%
\noindent
\section{Introduction}

%There has been 
New interest has emerged in jet noise modeling in last decade  after numerical simulations of jet turbulence provided some evidence that the well-known idea of `wave packet'-like structures (Crow $\&$ Champagne 1972) possibly embedded in the jet appear to play a direct role in low frequency sound radiation. 
This was discussed in several recent review papers for example, by Lele $\&$ Nichols (2014, p. 4-5), Suzuki (2013) and Jordan $\&$ Colonius (2013) and the references cited therein.
Our focus here, however, is on the alternative, acoustic analogy approach. 
In particular we consider the development of a robust mathematical model for the acoustic spectrum of an unheated round jet flow using recent developments we have made in the low frequency asymptotic analysis of the adjoint linearized Euler equations (ALEE).
The latter set of equations determine the so-called `propagator' tensor (often referred to as simply the `propagator') which appears as a convolution product with a generalized stress tensor (that encapsulates all flow unsteadiness effects) in the acoustic spectrum of the generalized acoustic analogy (Goldstein, 2003).
The propagator tensor plays an important role in the determining the correct low frequency roll-off in the predicted acoustic spectrum at frequencies upto the peak noise. 

Goldstein`s (2003) formulation provides the most comprehensive starting point for jet noise modeling under the set of approaches collectively referred to as {\it acoustic analogies} of the type first invented by Lighthill (1952).
All acoustic analogy models begin by re-arranging Navier Stokes equations so that the left hand side operator governs the wave propagation in some form in the same manner as the response of a linear system  forced by a non-linear source term on the right hand side (representing the turbulence localized within the jet) does. 
Hence, while various acoustic analogy models may differ in the interpretation of what terms constitute the wave propagation and the mathematical definition of the `sound source', physically, the turbulence-induced pressure fluctuations ($p^\prime$) are sustained (i.e. balanced) by the local transfer of momentum that occurs both randomly and chaotically in a region where the source term is non-zero and is a stationary random function of time.
The basic approximation thus boils down to assuming that the statistical structure (viz. the auto-covariance) of the source term is a known function that can be modeled appropriately; for example, by using a computational and/or experimental database (see Karabasov {\it et al}. 2010 and Lele {\it et al}. 2010).

As opposed to previous acoustic analogies (e.g. Lilley 1971),
% check Powell. Howe. Philips.
Goldstein's theory uses an ab initio decomposition of the  fluid-mechanical variables into their base flow and residual (defined relative to the base flow) components. The generalized analogy uses non-linear quantities as the dependent fluid mechanical variables to define the wave propagation operator (Eq. A.1 in Goldstein, 2003)  
%$L_{\mu\nu}$ 
and, most importantly, to allow the source term on the right hand side to be expressed in terms of the generalized Reynolds stress tensor, $e_{\lambda j}$, in a rather simple fashion (we define this term later).
%the right hand source terms to remain algebraically compact.
%
The use of non-linear variables in this way does not pose any technical difficulty in determining the resulting sound field however, because the non-linear pressure variable (Eq. 2.16 in Goldstein 2003) reduces to the ordinary acoustic pressure in the far field where the fluid is at rest and $p^\prime$ is governed by the homogeneous wave equation.
Moreover, since the linearized Euler equations possess a linear differential operator
%$L_{\mu\nu}({\boldsymbol y}, \tau)$ 
acting upon the residual component, the exact solution for the pressure fluctuation, $p^\prime$, at the observation point $({\boldsymbol x}, t)$, 
%(i.e. the 4th component of the dependent variable vector $u_\nu$) 
can be found by formally inverting that operator using Green's theorem (see 2.22 $\&$ App. A in Goldstein 2003) together with an adjoint vector Green's function, $g_{\nu4}^a({\boldsymbol y}, \tau | {\boldsymbol x},t)$, also defined later in $\S.$2.
The pressure perturbation, $p^\prime({\boldsymbol x}, t)$, is therefore given as a volume integral where the integrand is  a convolution product of a propagator tensor and a generalized stress tensor (that is linearly related to the fluctuating Reynolds stress  $e{}_{ij}^\prime$ in isothermal conditions) and whose auto-covariance, $R_{ijkl}$, is assumed known (as required to form an analogy).
%that, as mentioned, is a stationary random function of time.
%
When the base flow is taken as the {steady} jet mean flow (usually found via a Reynolds-averaged Navier-Stokes (RANS) calculation or the steady mean field of a Large-Eddy Simulation), the Fourier transform of the propagator is time-independent and is a function of the mean flow field and a vector Green's function of the adjoint linearized Euler equations (ALEEs).
This approach has proven to be successful for a number of test cases involving axisymmetric jets at a variety of acoustic Mach numbers and observation angles (Goldstein $\&$ Leib, 2008, hereafter referred to as G$\&$L). 
It has also shed light on what impact jet mean flows have on the far-field radiated sound for both heated and unheated conditions (Afsar {\it et al}., 2011 $\&$ 2019).
Any remaining issues then largely involve: (a). development of robust models for $R_{ijkl}$ and (b). determination of an appropriate solution to the adjoint vector Green`s function and, therefore, the propagator tensor. 

The present contribution focuses on the propagation aspect of the jet noise problem. We use the fact that non-parallel flow effects enter the lowest order asymptotic expansion of the propagator tensor when use is made of the low frequency asymptotic theory developed by Goldstein, Sescu $\&$ Afsar (2012, hereafter referred to as GSA) that appeared to capture the qualitative structure of non-parallelism found in the full numerical solution of the ALEEs (Karabasov {\it et al}., 2013). 
%
%%The importance of this work is clear: low frequency spectra constitute the peak jet noise observed at small observation angles;  mathematical models of the latter are useful for noise control strategies that seek to reduce the maximum radiated sound without need for long-time ALEE calculations to determine the adjoint vector Green's function.
%
That is, inclusion of mean flow spreading effects into the propagator solution can  increase the low-frequency radiation by almost 10 Decibels (dB) at $ \theta= 30^o$ on a high subsonic jet compared to the equivalent parallel flow solution of the ALEEs (Karabasov {\it et al}. 2010 and 2013).
GSA constructed an asymptotic solution to the adjoint vector Green's function to explain this finding by using a slowly diverging jet approximation in which jet spread rate, $\epsilon$, is asymptotically small, inasmuch as $\epsilon \ll O(1)$ where the propagator is sought at low frequencies of the same order as the jet spread rate (i.e. $\omega \sim \epsilon$) and is then matched with the outer wave equation solution at $O(1/\epsilon)$ radial distances.
Using this scaling, the dominant '1-2' propagator component that multiplies the '1-2' Reynolds stress in the acoustic spectrum formula (where $(1,2)$ refer to streamwise and transverse velocity fluctuations, respectively; see G$\&$L and Afsar, 2010) is everywhere different from the parallel flow result in the jet (and not just in the critical layer as in G$\&$L). 
The importance of this work is clear: low frequency sound is the main component of the peak jet noise at small observation angles where the sound field is maximum;  mathematical models of the latter are useful for noise control strategies that seek to reduce the maximum radiated sound without need for long-time ALEE calculations to determine the adjoint vector Green's function.

While GSA illustrated how the qualitative structure of the '1-2' propagator component based on this scaling differed from the parallel flow solution, Afsar {\it et al}. (2016) assessed its predictive capability using Reynolds-averaged Navier-Stokes (RANS) mean flow solutions to calculate the appropriate component of the adjoint vector Green's function and the relevant propagator term. However, they did not compare their turbulence model to LES or experiment.
%
%Their main numerical result (figure 5.3a) con- firms that an accurate prediction of the far-field sound can be made using this asymptotic approach. The predictions generally break down (i.e., rapidly decrease), however, above the peak Strouhal number (at St = 0.2), but that is not altogether unexpected owing to low-frequency applicability of the theory.
%
Our aim here is to investigate predictive capability of this asymptotic theory and, more broadly, to assess its limit of applicability in the parameter range of temporal frequency, acoustic Mach number and observation angle when the turbulence model is appropriately validated.
For the high subsonic jet, our noise predictions then extend to Strouhal numbers ($St$) beyond the peak frequency, i.e., near $St\sim 0.6$. We further extend this to more $O(1)$ frequencies by using an approximate composite Green's function and propagator within the acoustic spectrum formula
that gives much closer agreement over the entire frequency range for which acoustic data exists but necessarily introduces some empiricism into the model to estimate the transverse correlation length scale.  

We study two axisymmetric jets 
%in the paper %(one of which was not investigated in our previous work) 
with subsonic acoustic Mach numbers defined as $Ma=U_j/c_\infty$, where $U_j$ is the jet exhaust velocity and $c_\infty$ is the speed of sound at infinity. 
Under the Tanna matrix (1977; Bridges, 2006), these conditions are: SP07 ($Ma =0.9$ $\&$ $TR = 0.84$) and SP03 ($Ma =0.5$ $\&$ $TR = 0.95$), where $TR$ is the jet static temperature ratio. The jet total temperature is $1.0$ in both cases.

The paper begins by reviewing the GSA analysis using the simplified presentation 
%of the theory 
of Afsar {\it et al}. (2019). The mean flow  was obtained by the NASA {\sc{Wind-US}} code (Nelson $\&$ Power 2001; Nelson 2010) and the acoustic predictions obtained are discussed in  $\S$.(3).
%
%

%\vspace*{-10pt}
\section{Asymptotic analysis within the generalized acoustic analogy}
We fix ideas by considering a turbulent jet flow of $O(1)$ acoustic Mach number $Ma = U_J/c_\infty$ spreading downstream .
We let the (dimensional) pressure $p$, density $\rho$, enthalpy $h$, and speed of sound, $c$, satisfy the ideal gas law equation of state $p = \rho c^2/\gamma$, where $h= c^2/(\gamma-1)$
and $\gamma$ denotes the specific heat ratio. 
The acoustic spectrum at the observation point, ${\boldsymbol x} = (x_{_1}, {\boldsymbol x}_{_T}) = (x_{_1},x_{_2},x_{_3})$, is given by the Fourier transform 
%\vspace*{-10.0pt}
\begin{equation}
\label{eq:Iom}
I({\boldsymbol x}, \omega) 
\equiv
\frac{1}{2 \pi}
\int\limits_{-\infty}^{\infty}
e^{i \omega\tau}
\overline{
p^\prime({\boldsymbol x}, t)
p^\prime({\boldsymbol x}, t+\tau)
}
\,d\tau,
\end{equation}
%
%\vspace*{-0.1pt}
%
of the far-field pressure auto-covariance, $\overline{
p^\prime({\boldsymbol x}, t)
p^\prime({\boldsymbol x}, t+\tau)
}$. The former is also given by a volume integral over a unit volume of turbulence at ${\boldsymbol y} = (y_{_1}, {\boldsymbol y}_{_T}) = (y_{_1},y_{_2},y_{_3})$ in the jet as %
\begin{equation}
\label{eq:Iom2}
I({\boldsymbol x}; \omega) 
=
\int
\limits_{V_\infty({\boldsymbol y})}
I({\boldsymbol x}, {\boldsymbol y};\omega) 
\,d{\boldsymbol y},
\end{equation}
%({\bm y}, \tau)
where, $V_\infty({\boldsymbol y})$ is the entire source region, $p^\prime ({\boldsymbol y}, \tau) \equiv p({\boldsymbol y}, \tau) -\bar{p}({\boldsymbol y})$ and over-bars are being used to denote time averages defined by:
\begin{equation}
\label{eq:t_avg}
\bar{\bullet}({\boldsymbol x}) 
\equiv
\lim_{T\rightarrow\infty}
\frac{1}{2T}
\int\limits_{-T}^{T}
\bullet({\boldsymbol x}, t)
\,dt,
\end{equation}
where ${\bullet}$ in (\ref{eq:t_avg}) is a place holder for any fluid mechanical variable.

G $\&$ L showed that the integrand on right side of (\ref{eq:Iom2}) is given by the exact integral solution,
\begin{equation}
\label{eq:Iom3}
I({\boldsymbol x}, {\boldsymbol y};\omega) 
=
(2\pi)^2
\Gamma_{\lambda, j}
({\boldsymbol y}| {\boldsymbol x}; \omega)
\int
\limits_{V_\infty({\boldsymbol \eta})}
\Gamma{}_{\mu, l}^*
({\boldsymbol y} + {\boldsymbol \eta}| {\boldsymbol x}; \omega)
\mathcal{H}_{\lambda j \mu l}
({\boldsymbol y}, {\boldsymbol \eta}; \omega)
\,d{\boldsymbol \eta}.
\end{equation}
Asterisks denote complex conjugate and the Einstein summation convention is being used with the Greek tensor suffixes ranging $(\lambda,\mu)=(1,2,3,4)$ and the Latin suffixes $(i,j,k,l)=(1,2,3)$.
The mean flow now enters through the Fourier transformed propagator tensor
\begin{equation}
\label{eq:Prop}
\Gamma_{\lambda, j}
({\boldsymbol y}| {\boldsymbol x}; \omega)
\equiv
\Lambda_{\lambda\sigma,j}
({\boldsymbol y})
G_\sigma
({\boldsymbol y}| {\boldsymbol x}; \omega)
:=
\left(
\delta_{\lambda\sigma}
\frac{\partial}{\partial y_j}
-
(\gamma-1)
\delta_{4\sigma}
\frac{\partial\tilde{v}_\lambda}{\partial y_j}
\right)
G_\sigma
({\boldsymbol y}| {\boldsymbol x}; \omega)
\end{equation}
that involves an inner tensor product in suffix $\sigma$, of operator $\Lambda_{\lambda\sigma,j}
({\boldsymbol y})$, that spans $(4\times4\times3)$ dimensions corresponding to suffixes $(\lambda,\sigma,j)$ where comma after $j$ indicates that this suffix belongs to a derivative, and the first four components of the Fourier transform
\begin{equation}
\label{eq:GFT}
{ \boldsymbol{G}}
({\boldsymbol y}| {\boldsymbol x}; \omega)
=
\frac{1}{2 \pi}
\int\limits_{-\infty}^{\infty}
e^{i \omega(t-\tau)}
{\boldsymbol g}{}_{4}^{a}
({\boldsymbol y}, t - \tau| {\boldsymbol x})
\,d(t-\tau)
%\hspace{.5cm}
%\sigma=(1,2,3,4,5)
,
\end{equation}
of the five-dimensional adjoint vector Green's function, ${\boldsymbol g}{}_{4}^{a}
({\boldsymbol y}, \tau| {\boldsymbol x}, t)$, 
%(where $\sigma=(1,2,3,4,5)$) 
that appears on the left hand sides of the five ALEE given by  (2.9a) $\&$ (2.13) in Goldstein 2003, and (3.1)--(3.3) of G$\&$L) subject to the strict causality condition
${g}{}_{44}^{a}
({\boldsymbol y}, t - \tau| {\boldsymbol x})=0$ for 
$t<\tau$ when $|{\boldsymbol x}|\rightarrow \infty$. 
As frequently commented in previous papers (Goldstein 2003, G $\&$ L and Leib $\&$ Goldstein, 2011), (\ref{eq:Iom3}) and (\ref{eq:Prop}) are completely general and apply to any localized turbulent flow, even in the presence of fixed solid surfaces whose boundaries are given by level curves $S({\boldsymbol y}) =  const.$ as long as ${\boldsymbol g}{}_{4}^{a}
({\boldsymbol y}, \tau| {\boldsymbol x}, t)$ is assumed to satisfy appropriate surface rigidity conditions $ \hat{\boldsymbol{n}}.{\boldsymbol g}{}_{4}^{a}
({\boldsymbol y}, \tau| {\boldsymbol x}, t) = 0$ where $\hat{\boldsymbol{n}}=\{\hat{n}_1,\hat{n}_2,\hat{n}_3\}$ denotes the unit normal to $S({\boldsymbol y})$.

In (\ref{eq:Prop}), $\delta_{\lambda\sigma}$ is the symmetric four-dimensional Kronecker delta (unit) tensor and tilde refers to the Favre averaged quantity $\tilde{\bullet} = \overline{\rho \bullet}/\bar{\rho}$, so that the four-dimensional mean velocity vector in (\ref{eq:Prop}) is $\tilde{v}_\lambda = \{\tilde{v}_i,0\}, i=(1,2,3)$.
The 5th component of ${ \boldsymbol{G}}
({\boldsymbol y}| {\boldsymbol x}; \omega)$ -- the Fourier transform of the adjoint Green's function for the continuity equation in the linearized Euler equations derived in Goldstein (2003, Eq. $2.9$a) -- does not enter the formula for the propagator, (\ref{eq:Prop}); it does, however, affect its solution through the ALEE (see (4.8)-(4.10) of G $\&$ L) given by:
\begin{subequations}
\label{eq:GAA}
\begin{align}
-{D}_0 G_i + G_j \frac{\partial \tilde{v}_j}{\partial y_i}
-\widetilde{c^2}\frac{\partial G_4}{\partial y_i} +(\gamma-1)\tilde{X}_i G_4
-\frac{\partial G_5}{\partial y_i} & = 0 \\
-{D}_0 G_4 - \frac{\partial G_i}{\partial y_i}
+(\gamma-1)G_4 \frac{\partial\tilde{v}_i}{\partial y_i} & = \frac{\delta({\boldsymbol{x}-\boldsymbol{y}})}{2\pi} \\
-{D}_0 G_5 +\tilde{X}_i G_i & = 0, 
%\\
%
%y & = ax^{3}+bx^{2}+cx+d
\end{align}
\end{subequations}
where ${D}_0 \equiv i\omega + \tilde{{\boldsymbol v}}({\boldsymbol y}).{\boldsymbol{\nabla}}$ is the convective derivative and  ${\boldsymbol\nabla} \equiv \{\partial/\partial y_{_1}, \partial/\partial y_{_2}, \partial/\partial y_{_3}\}$ is the three-dimensional gradient operator.

Reciprocity (see pp. 878--886 of Morse and Feshbach, 1953) of the space-time Green's function demands that ${\boldsymbol g}{}_{4}^{a}
({\boldsymbol y}, \tau| {\boldsymbol x}, t)={\boldsymbol g}{}_{4}
({\boldsymbol x}, t|{\boldsymbol y}, \tau) $ and therefore (after taking temporal Fourier transforms) that the independent variable ${\boldsymbol y}$ in (\ref{eq:GAA}) corresponds to the actual physical source point where, ${\boldsymbol x}$, is the observation point, which is taken as a parameter in the solution and located in the far field, $|{\boldsymbol x}|\rightarrow\infty$.
%T
The coefficients in (\ref{eq:GAA}) depend on the mean flow field through $\tilde{v}_i = (\tilde{v}_1, \tilde{v}_2, \tilde{v}_3)$;  $\widetilde{c^2}({\boldsymbol y})\equiv {\gamma \bar{p}}/{\bar{\rho}}$, the mean flow speed of sound squared, and 
\begin{equation}
\label{Xdef}
\boldsymbol{\tilde{X}}({\boldsymbol y}) = (\boldsymbol{\tilde{v}}.\boldsymbol{\nabla}) \boldsymbol{\tilde{v}},
\end{equation}
is the mean flow advection vector.

The tensor $\mathcal{H}_{\lambda j \mu l}
({\boldsymbol y}, {\boldsymbol \eta}; \omega)$ in the acoustic spectrum formula, (\ref{eq:Iom3}), is related to the Fourier transform 
 \begin{equation}
\label{eq:HFT}
{H}_{\lambda j \mu l}
({\boldsymbol y}, {\boldsymbol \eta}; \omega)
=
\frac{1}{2 \pi}
\int\limits_{-\infty}^{\infty}
e^{i \omega\tau}
{R}_{\lambda j \mu l}
({\boldsymbol y}, {\boldsymbol \eta}; \tau)
\,d(\tau)
\end{equation}
of the generalized auto-covariance tensor,
\begin{equation}
\label{eq:Rijkl}
{R}_{\lambda j \mu l}
({\boldsymbol y}, {\boldsymbol \eta}; \tau)
\equiv
\lim_{T\rightarrow\infty}
\frac{1}{2T}
\int\limits_{-T}^{T}
%\left[
%\rho
%v{}_\lambda^\prime
%v{}_j^\prime
%-
%\overline{\rho
%v{}_\lambda^\prime
%v{}_j^\prime}
%%\right]
e{}_{\lambda j}
({\boldsymbol y}, \tau)
%javascript:void(0);
e{}_{\mu l}
({\boldsymbol y} + {\boldsymbol \eta}, \tau+\tau_0)
\,d\tau_0,
\end{equation}
of the stationary random function, $e{}_{\lambda j}({\boldsymbol y}, \tau)= [\rho v{}_\lambda^\prime v{}_j^\prime -
\overline{\rho
v{}_\lambda^\prime
v{}_j^\prime}
]({\boldsymbol y}, \tau)
$, by the linear transformation $\mathcal{H}_{\lambda j \mu l}({\boldsymbol y}, {\boldsymbol \eta}; \omega):=\epsilon_{\lambda j \sigma m} H_{\sigma m \gamma n} ({\boldsymbol y}, {\boldsymbol \eta}; \omega) \epsilon_{\mu l \gamma n}$.
Comparing (5.12) to (5.13) in G $\&$ L (2008) and using appropriate outer products of unit tensors (see also sentence below (\ref{eq:GFT})) in suffixes $(\lambda,j,\sigma, m)$ allows definition of the tensor as, $\epsilon_{\lambda j \sigma m} \equiv \delta_{\lambda\sigma} \delta_{j m} - \delta_{\lambda j} \delta_{\sigma m}(\gamma-1)/2$ in the linear relation for $\mathcal{H}_{\lambda j \mu l}$ above.
The four-dimensional perturbation velocity, $v{}_\lambda^\prime ({\boldsymbol y},\tau) \equiv v{}_\lambda ({\boldsymbol y},\tau)- \tilde{v}{}_\lambda ({\boldsymbol y})$ in which $v{}_\lambda^\prime =v{}_i^\prime$ is the ordinary fluid velocity perturbation when suffix, $\lambda=i= (1,2,3)$, otherwise $v{}_\lambda^\prime =v{}_4^\prime$ is proportional to enthalpy fluctuation (discussed further in Afsar {\it et al.} 2019).

GSA derived an asymptotic model for the Fourier transformed propagator, $\Gamma_{\lambda, j}$, for a slowly diverging jet flow at temporal frequencies of the order of the small jet spread rate, that is, $\omega = O(\epsilon)$.
The lowest order inner equations in GSA's analysis (Eqs. 5.18--5.20) reduced to single second-order hyperbolic PDE for a composite Green's function variable when the independent variables were transformed using the streamwise mean flow component, $U$, as one of the independent variables. 
However, in recent work, Afsar {\it et al}. (2019) found that this transformation can easily be applied to the Fourier transformed ALEE (\ref{eq:GAA}a--c) {\it at the outset}, prior to any asymptotic analysis (in other words at $\omega = O(1)$ frequencies). The advantage of which is that when the latter is used, in the form of method of multiple scales and matched asymptotic expansions (in that order), the basic inner equation immediately follows. 

\subsection{Summary of Afsar {\it et al.}'s (2019) reformulation of the GSA theory}

\subsubsection{Transformation of (\ref{eq:GAA}) at $O(1)$ spread rates}
Let independent variables $({\boldsymbol y}, \tau)$ be normalized by $O(1)$ characteristic length $D_J$ and time $D_J/U_J$, respectively where $U_J$ $\&$  $D_J$ are the mean velocity and nozzle exit diameter respectively. The dependent variables in the ALEEs (\ref{eq:GAA}a--c), $(\tilde{{\boldsymbol v}},p,\rho)$, may then be normalized by $U_J$, $\rho_J U_J^2$ and $\rho_J$. When $({\boldsymbol e}_1, {\boldsymbol e}_r, {\boldsymbol e}_\phi)$ is orthogonal basis vectors in a cylindrical co-ordinate space, ${\boldsymbol G} = (G_1, G_r, G_\phi)$ in (\ref{eq:GAA}a-c) can be expressed as a linear function of that basis 
%(Temple 1960, p.8) 
by $(G_i {\boldsymbol e}_i){\boldsymbol e}_j = G_1 \delta_{j1} + G_r \delta_{jr} + G_\phi \delta_{j\phi}$. 
%where ${\boldsymbol G} = (G_1, G_r, G_\phi)$ are its respective components of ${\boldsymbol G}$ in the basis %$({\boldsymbol e}_1, {\boldsymbol e}_r, {\boldsymbol e}_\phi)$ 
The mean flow field, commensurate with an axisymmetric jet, has components, ${\boldsymbol v} = (U, V_r)$ where (at this point) we leave the jet spread rate arbitrary at $\epsilon=O(1)$.

Following GSA we take $U$ to be one of the independent variables of choice; i.e., $(y_1,r)$ $\rightarrow(y_1,U)$ where $r \equiv |{\boldsymbol y}_T|= \sqrt{y{}_2^2 + y{}_3^2}$. The co-ordinate surfaces $U(y_1,r) = const.$ and $y_1 = const.$ are such that ${\boldsymbol\nabla} U. {\boldsymbol\nabla} y_1 = 0$ at any fixed radial location $r$. Since the gradient operator
%$\nabla_{\boldsymbol y}\equiv {\boldsymbol e}_1 \partial/\partial y_1 + {\boldsymbol e}_r \partial/\partial r$, 
shows that ${\boldsymbol e}_1 \equiv { \boldsymbol\nabla} y_1$ and ${\boldsymbol\nabla} U \equiv {\boldsymbol e}_1 {\partial U}/{\partial y_1} + {\boldsymbol e}_r {\partial U}/{\partial r}$, the choice of independent variables implies that ${\boldsymbol \nabla} U.{\boldsymbol \nabla} y_1 = {\partial U}/{\partial y_1 }=0$ in the transformed co-ordinate system. Using the fact that 
${ \boldsymbol G}
(y_1, r, \phi| {\boldsymbol x}; \omega)$ is implicitly related to $\tilde{G}_i=\tilde{G}_i (y_1, U,\phi|{\boldsymbol x}; \omega)$ via $\tilde{G}_i
(y_1, U(y_1,r),\phi|{\boldsymbol x}; \omega)
=  
{G}{}_i
(y_1,r,\phi|{\boldsymbol x}; \omega)$
%
%\begin{equation}
%\label{G_implicit}
%\tilde{G}_i
%(y_1, U(y_1,r),\phi|{\boldsymbol x}; \omega)
%=  
%{G}{}_i
%(y_1,r,\phi|{\boldsymbol x}; \omega),
%\end{equation}
%
the orthogonality condition and the Chain rule in $(y_1,U)$ co-ordinates similarly shows that the mean flow advection vector $\tilde{X}_i=(\tilde{X}_1, \tilde{X}_r)$ (in \ref{eq:GAA}a-c) takes the more general form than that given by Eq. (5.15) in GSA.
Moreover, operator $D_0$ acting on ${ \boldsymbol{G}}
(y_1, r, \phi| {\boldsymbol x}; \omega)$ in Eqs. (\ref{eq:GAA}a--c) may be transformed to
\begin{equation}
\label{D0_trans}
% \begin{split}
D_0
{G}_i
(y_1,r)
=  
  \left( i\omega + U\frac{\partial}{\partial y_1} + V_r \frac{\partial}{\partial r}
  \right) {G}{}_i
\equiv
\left(
\tilde{D}_0 + \tilde{X}_1 \frac{\partial}{\partial U}
\right)
\tilde{G}_i (y_1, U),
 % \end{split}
   \end{equation}
where we have suppressed the remaining arguments in $G_i$, $\tilde{D}_0 \equiv i\omega + U{\partial}/{\partial y_1}$ and $\tilde{\boldsymbol{X}}$ is given by (\ref{Xdef}).
Since ${\partial U}/{\partial r} = ({\partial r}/{\partial U})^{-1}$
and the Chain rule shows that ${\partial}/{\partial U} = ({\partial r}/{\partial U})/{\partial}/{\partial r}$, the $i=r$ component of (\ref{eq:GAA}a) is transformed to the following result
\begin{equation}
\label{G_1}
\tilde{G}_1
(y_1, U)
%(y_1, U,\psi|{\boldsymbol x}; \omega)
=  
\widetilde{c^2}
\frac{\partial \tilde{G}_4} {\partial U}
+
\frac{\partial \tilde{G}_5} {\partial U}
+
\tilde{S}_r
(y_1, U)
\end{equation}
that generalizes (5.23) of GSA to jets for which $\epsilon = O(1)$. The right hand side term is discussed below and acts to couple the various components of the ALEE (\ref{eq:GAA}); it is one component of the vector  $\tilde{S}_i = \{\tilde{S}_1,\tilde{S}_r,\tilde{S}_5\}$. 

Inserting the second member of (\ref{D0_trans}) $\&$ (\ref{G_1}) into (\ref{eq:GAA}c) shows that it can be transformed to $\tilde{D}_0 \tilde{\nu}
(y_1, U)
=  
\widetilde{c^2}
D_0 \tilde{G}_4
%(y_1, U)
+
\tilde{S}_5
 (y_1, U)$
for the Green's function variable, $\tilde{\nu}= \tilde{\nu}
(y_1, U) \equiv \widetilde{c^2} \tilde{G}_4 + \tilde{G}_5$ when $\widetilde{c^2} = f (U )$ in which $f$ can be an arbitrary function but will be specified shortly to eliminate any `$\tilde{G}_4$ terms' appearing on the left hand side of (\ref{hyb_eqn1}).
%(\ref{nu_eqn}).
%
To set about doing this we first integrate (\ref{G_1}) by parts to re-write its right hand side in terms of $\tilde{\nu}(y_1, U)$ and insert the result, (\ref{D0_trans}) $\&$ the relation above for $\tilde{D}_0 \tilde{\nu}
(y_1, U)$ into the $i=1$ component of (\ref{eq:GAA}a) to give:
\begin{equation}
\label{hyb_eqn1}
\frac{\partial }{\partial U}
\tilde{D}_0 \tilde{\nu}
-
\frac{1}{\widetilde{c^2}}
\frac{\partial \widetilde{c^2}}{\partial U}
\tilde{D}_0 \tilde{\nu}
+
\tilde{X}_1
\frac{\partial^2 \tilde{\nu}}{\partial U^2}
-
%&
\tilde{X}_1
%
%\left[
%\textcolor{red}{
\left[
(\gamma-1)
+
\frac{\partial^2 \widetilde{c^2}}{\partial U^2}
\right]
%}
%\right]
\tilde{G}_4 
=
-\tilde{S}_1 
%\frac{\tilde{S}_5}{\widetilde{c^2}}
%(y_1, U)
+
\left(
\frac{\tilde{S}_5}{\widetilde{c^2}}
+
D_0
\tilde{S}_r
\right). 
%(y_1, U)
\end{equation}

Afsar {\it et al}. (2019) then show that the term in square brackets in  Eq.(\ref{hyb_eqn1}) is identically zero when the jet is isothermal and $\widetilde{c^2}$ assumed to satisfy the Crocco relation (inasmuch as $\widetilde{c^2}(U)  = c{}^2_\infty - (\gamma-1)U^2/2$, where $c{}_\infty$ is the speed of sound at infinity) or heated and, therefore, satisfies the Crocco-Busemann relation. 
Hence, integrating by parts in (\ref{hyb_eqn1}), shows that the combined variable, $\tilde{\nu}(y_1,U)$, is determined by the following partial differential equation (PDE):
\begin{equation}
\label{Hyp2}
\mathcal{L}
%{L}
%(y_1, U(y_1, r))
\tilde{\nu}
(y_1, U)
=
\mathcal{F}(\tilde{\boldsymbol S}),
\hspace{0.5cm}
\textnormal{for}
\hspace{0.25cm}
\epsilon = O(1).
\end{equation}
Eq. (\ref{Hyp2}), which replaces (\ref{eq:GAA}a)--(\ref{eq:GAA}c), also generalizes (5.30) $\&$ (5.31) in GSA where
\begin{equation}
\label{L_Hyp}
\mathcal{L}(y_1, U)
\equiv
\widetilde{c^2}
\frac{\partial }{\partial U}
%\left(
\frac{1}{\widetilde{c^2}}
\tilde{D}_0 
%\tilde{\nu}
%\right)
+
\tilde{X}_1
\frac{\partial^2 }{\partial U^2},
\end{equation}
is a hyperbolic operator but now for the arbitrary axisymmetric jet mean flow field ${\boldsymbol v}(y_1, r) = (U, V_r)$ at $O(1)$ jet spread rates.
The right hand side of (\ref{Hyp2}) is 
the functional $\mathcal{F}(\tilde{\boldsymbol S}) = (\delta_{i1} + \delta_{ir}D_0 -\delta_{ir}/\widetilde{c^2})S_i $ where $\tilde{\boldsymbol S}(y_1, U)= \{\tilde{ S}_1,\tilde{S}_r, \tilde{S}_5\}(y_1, U)$ is linearly related to the adjoint Green's function component for the radial momentum equation, $\tilde{G}_r$, and the 
%%higher order 
mean flow component, $V_r$, 
%%(cf. \ref{eq:Meanflow_exp}) 
via: $\tilde{S}_1
(y_1, U)
=  
({\partial V_r}/{\partial y_1})
\tilde{G}_r (y_1, U)$ and $\tilde{S}_5
 (y_1, U)
=
\tilde{X}_r
\tilde{G}_r
+\tilde{X}_1
\tilde{S}_r$
where
\begin{equation}
\label{S_r}
\tilde{S}_r
(y_1, U)
(y_1, U,\psi|{\boldsymbol x}; \omega)
=  
\frac{\partial r} {\partial U}
\left[
\left(
D_0
-
\frac{\partial {V}_r} {\partial r}
\right)
\tilde{G}_r
(y_1, U)
-(\gamma-1)
\tilde{X}_r
\tilde{G}_4
(y_1, U)
\right]
(y_1, U)
\end{equation}
Equation (\ref{Hyp2}) is simply a direct re-arrangement of Fourier transformed ALEE, (\ref{eq:GAA}a-c) where $\mathcal{F}(\tilde{\boldsymbol S} )$ is defined explicitly in Afsar {\it et al}. (2019).
Although it is valid for an arbitrary axisymmetric jet flow with mean flow components, ${\boldsymbol v} = (U, V_r)$, where the speed of sound is determined by Crocco relation in isothermal flows that are of interest in this paper and in which %\newline  
$\tilde{G}_\mu=\tilde{G}_\sigma (y_1, U,\phi|{\boldsymbol x}; \omega)$ is the appropriate $O(1)$ frequency adjoint vector Green`s function solution ($\sigma = 1,2,...5$), it is just as complex as the original ALEEs in (\ref{eq:GAA}). 
This is because $\mathcal{F}(\tilde{\boldsymbol S})$ depends on the `leftover terms', $\tilde{S}_i = \{\tilde{S}_1,\tilde{S}_r,\tilde{S}_5\}$, on the right hand side of (\ref{Hyp2}) which transform it to a mixed PDE that requires the solution of $4$ coupled equations for $(\tilde{\nu},\tilde{G}_4,\tilde{G}_r, \tilde{G}_\phi)$ using the $\tilde{D}_0 \tilde{\nu}
(y_1, U)$ relation above (\ref{hyb_eqn1}), (\ref{Hyp2}) and $i=(r,\phi)$ components of (\ref{eq:GAA}a) when (\ref{G_1}) is substituted for $\tilde{G}_1$.
However, Afsar {\it et al}. (2019) show that the right hand side of (\ref{Hyp2}) remains exactly at $o(1)$ in the small jet spread rate limit ($\epsilon\ll O(1)$) when the temporal frequency is appropriately re-scaled. Therefore, $\mathcal{F}(\boldsymbol{S})$ remains asymptotically sub-dominant in this limit. We summarise this next and show it leads to an asymptotic expansion of $\Gamma_{\lambda, j}$ that at its lowest order involves only a single term.

\subsubsection{Elimination of $\tilde{\boldsymbol S}(y_1, U)$ in (\ref{Hyp2}) at lowest order in $\epsilon$}
\label{sec:2b}
%({\boldsymbol y}| {\boldsymbol x}; \omega)$}
%
That an axi- symmetric mean flow diverges with an asymptotically small spread rate, $\epsilon \ll O(1)$, is consistent with experiments by Panchapasekan $\&$ Lumley (1993) which indicate (see p.101ff. in Pope 2000) that $\epsilon$ is virtually constant with Reynolds number and nearly equal to $0.1$ at isothermal conditions.
We therefore take the mean flow to vary over a slow streamwise length, $Y \equiv\epsilon y_1= O(1)$, corresponding to long streamwise length scales $y_1$, relative to an origin placed at the nozzle exit plane. Whence, it must expand according to (A.1--A.2) in G $\&$ L; viz.
\begin{equation}
\label{eq:Meanflow_exp}
%\begin{align}
%
{\tilde{v}_i}=\{U(Y), V_r(Y, U)\}  =\begin{cases}
      U +\epsilon U^{(1)}(Y, U)+ O(\epsilon^2),\hspace{0.75cm} i=1 \\
      \epsilon (V_r + \epsilon V_r^{(2)})(Y, U) + O(\epsilon^3), \hspace{0.25cm} i=r 
            \end{cases} 
%
%\end{align}
\end{equation}
when $\widetilde{c^2}$ is determined by the Crocco relation (below \ref{hyb_eqn1}).
We have not put superscripts on the lowest order mean flow components, that would otherwise appear as $(U^{(0)},V{}_r^{(1)})$ respectively; they will be taken as that computed by the RANS solution.
Moreover at this order in $\epsilon$: ${\bar \rho}(Y, U) =  \rho(U)$ and ${\bar p}(Y, U) = const.$ and the mean flow advection vector, $X_i(\boldsymbol{y})$, that enters in $\tilde{S}_i = \{\tilde{S}_1,\tilde{S}_r,\tilde{S}_5\}$, similarly expands as
\begin{equation}
\label{eq:X_exp}
{\tilde{X}_i}=\{\tilde{X}_1, \tilde{X}_r\}(Y, U)  =\begin{cases}
      \epsilon \bar{X}_1(Y, U) +\epsilon^2 \tilde{X}{}_1^{(2)}(Y, U)+ O(\epsilon^3),\hspace{0.25cm} i=1 \\
      \epsilon^2\bar{X}{}_r^{(2)}(Y, U) + O(\epsilon^3), \hspace{2.2cm} i=r 
            \end{cases} 
\end{equation}
where the leading streamwise term, $\bar{X}{}_1^{(1)}\equiv \bar{X}{}_1 = V_r (\partial U/\partial r)$ and $\bar{X}{}_r^{(2)}=(U\partial/\partial Y + V_r \partial/\partial r) V_r$.
Hence, measured from the jet centerline, the mean flow separates into an inner region, given by (\ref{eq:Meanflow_exp}) $\&$ (\ref{eq:X_exp}), where (inner) radial co-ordinate $r = O(1)$, and an outer region where this expansion break downs; i.e., at large radial locations (with respect to inner variable, $r$) for which $R \equiv \epsilon r = O(1)$. 

But the long $O(1/\epsilon)$ streamwise variation of non-parallel flow alters the leading order structure of propagator, ${\Gamma}_{\lambda, j}({\boldsymbol y}| {\boldsymbol x}; \omega)$, everywhere in the flow at $O(1)$ acoustic Mach numbers when ${\boldsymbol g}{}_{4}^{a}
({\boldsymbol y}, \tau| {\boldsymbol x}, t)$ modulates in time under an appropriate slow-time asymptotic scaling. 
In other words, the lowest order solution to ${\Gamma}_{\lambda, j}({\boldsymbol y}| {\boldsymbol x}; \omega)$ is governed by a non Rayleigh-type (or, in the time domain, a Lilley-Goldstein) equation in the inner region at  $O(1/\epsilon)$ acoustic wavelengths 
%i.e. this occurs at low frequencies where time variations are also slow and of the same order as the streamwise variations in the mean flow.
%
Hence ${\boldsymbol g}{}_{4}^{a}
({\boldsymbol y}, \tau| {\boldsymbol x}, t)$ depends on $\tau$ through re-scaled $O(1)$ time variable $\tilde{T}\equiv\epsilon\tau = O(1)$ 
%(cf. Wu $\&$ Huerre's ``slowly breathing'' modes)
inasmuch as the Strouhal number, $St$ (the scaled frequency), is of the order of the jet spread rate, $\epsilon$, in the solution to ${ \boldsymbol{G}}
({\boldsymbol y}| {\boldsymbol x}; \omega)$. 
The distinguished asymptotic scaling in this latter solution occurs when the $\epsilon \rightarrow 0$ limit is taken and the scaled frequency, $\Omega \equiv \omega/\epsilon = O(1)$ is held fixed. 
It is only at this limit, where the solution to the ALEE, (\ref{eq:GAA}a)--(\ref{eq:GAA}c), for ${ \boldsymbol{G}}
({\boldsymbol y}| {\boldsymbol x}; \omega)$ becomes asymptotically disparate as $\epsilon \rightarrow 0$ and -- like (\ref{eq:Meanflow_exp}) $\&$ (\ref{eq:X_exp}) -- divides into an inner solution where $r = O(1)$ and an outer solution valid at $R \equiv \epsilon r = O(1)$ distances from the jet axis. 
Similarly, at this limit the propagator, ${\Gamma}_{\lambda, j}({\boldsymbol y}| {\boldsymbol x}; \omega)$, is also everywhere different from the locally parallel flow.

Re-scaling the frequency $\omega = \epsilon\Omega$ and streamwise co-ordinate $y_1 = Y/\epsilon$ of the operator $D_0$ in (\ref{D0_trans}) shows that the latter operator acting on $\tilde{\nu}(Y, U)$ is given by,
\begin{equation}
\label{D0nu_trans}
% \begin{split}
D_0
\tilde{\nu}
(y_1,U)
=  
\epsilon  \left( i\Omega + U\frac{\partial}{\partial Y} + V_r \frac{\partial}{\partial r}
  \right) \tilde{\nu}
\equiv
\epsilon
\left(
\bar{D}_0 + \bar{X}_1 \frac{\partial}{\partial U}
\right)
\tilde{\nu}(Y, U),
 % \end{split}
   \end{equation}
where $\bar{D}_0 \equiv i\Omega + U{\partial}/{\partial Y}$. 
%at $\Omega=O(1)$
%
Eq. (\ref{D0nu_trans}) shows that $D_0 \tilde{\nu}=O(\epsilon)$ when $\tilde{\nu}(Y,U))$ expands with $O(1)$ term, which it must since the solution to $\tilde{\nu} (Y, U)$ in the outer region (see Eq.(5.40) and discussion at bottom of p.19 of GSA) expands in this manner. 
Afsar {\it et al.} (2019) note that  $\mathcal{F}(\tilde{\boldsymbol S})$ will then expand at least as $O(\epsilon^2)$ because, $\tilde{S}_i =\{\tilde{S}_1,\tilde{S}_r,\tilde{S}_5\}$ expands as $O(\epsilon^2)$ using (\ref{S_r}), (\ref{eq:Meanflow_exp}) $\&$ (\ref{D0nu_trans}), when the Green's function components $\tilde{G}_{(r,\phi)}$ expand as $O(1)$.

Although this would cause  on the right hand side of Eq. (\ref{Hyp2}), to drop out of the lowest order $\tilde{\nu}$--equation, this does not turn out to give the richest possible balance for $\tilde{G}_{(r,\phi)}$ in (\ref{eq:GAA}).
GSA show that the latter occurs when $\tilde{G}_{(r,\phi)}$ expands like $O(1/\epsilon)$ at lowest order in (\ref{eq:GAA}). $\mathcal{F}(\tilde{\boldsymbol S})$ still drops out of (\ref{Hyp2}) because $\tilde{G}_{(r,\phi)}$ must remain bounded on the jet axis.
By considering the conditions across the surface $r=0$ in the $i=\phi$ component of Eq. (\ref{eq:GAA}a) and using $\boldsymbol{\nabla}.\tilde{{\boldsymbol v}} \sim D_0\tilde{G}_4 = O(\epsilon)$ in the adjoint energy equation, Eq. (\ref{eq:GAA}b), 
%(in which $\boldsymbol{\nabla}.{\boldsymbol\tilde{v}} \sim D_0\tilde{G}_4 = O(\epsilon)$), 
it is easy to show that $\tilde{G}_{(r,\phi)}= 0$ at lowest order in Eq.(\ref{eq:GAA}) (see Afsar {\it et al.}, 2019, for more details). 
%
%$i=\phi$ component of (\ref{eq:GAA}a) and the adjoint energy equation (\ref{eq:GAA}b) (in which $\boldsymbol{\nabla}.{\boldsymbol\tilde{v}} \sim D_0\tilde{G}_4 = O(\epsilon)$), shows that $\tilde{G}_{(r,\phi)}= 0$ at lowest order in (\ref{eq:GAA}), (\ref{Hyp2}), (\ref{S_r}) $\&$ the relations defined above it. 

The final simplification to the analysis comes as a consequence of using $\tilde{G}_{\phi}= 0$ in the $i=\phi$ component of (\ref{eq:GAA}a), which recovers the fact that the solution to (\ref{Hyp2}) is independent of azimuthal angle $\phi$.  
In other words, the Fourier transform of $\tilde{\nu} (Y,U, \phi|X, \Phi;\Omega)$ in the difference, $(\Phi-\phi)$, is given by
 \begin{equation}
%  \begin{split}
\label{nu_FT_azim}
\hat{\nu}{}^{ (n)} 
(Y, U) 
%(Y, U| X,|\boldsymbol{x}_T|,0 ; \Omega) 
=
 %&
\frac{1}{2 \pi}
\int\limits_{-\infty}^{\infty}
\tilde{\nu} (Y, U| X,|\boldsymbol{x}_T|,\Phi-\phi; \Omega) 
e^{i n(\Phi-\phi)}
\,d(\Phi-\phi) 
%\\
 \equiv 
%&
\delta(n)
\tilde{\nu} (Y, U)\mid_{(\Phi- \phi)=0}, 
%\tilde{\nu} (Y, U| X,|\boldsymbol{x}_T|,0; \Omega) 
% \end{split}
\end{equation}
%, (\phi- \Phi)=0
where $\delta(\bullet)$ is the Dirac delta function of argument $(\bullet)$. 
%and we have suppressed repeated variable list in the $\nu$--solution
Using (\ref{eq:GFT}), the solution, $\bar{\nu} (Y,U)$, is therefore given by the scaled Fourier transform (note error in pre-factor of Eq.5.8 in GSA):
 \begin{equation}
 \begin{split}
\label{eq:Scaled_G}
\tilde{\nu}(Y,U)
\equiv 
&
\frac{\epsilon}{4\pi c{}_\infty^2 |{\boldsymbol x}|}
e^{i\Omega X/c_\infty}
\bar{\nu}
(Y, U| X,|\boldsymbol{x}_T|,0 ; \Omega) 
\\
= &
\frac{1}{2 \pi\epsilon}
\int\limits_{-\infty}^{\infty}
e^{i \Omega(\tilde{T}_0-\tilde{T})}
(\widetilde{c^2} \tilde{g}{}_{4 4} + \tilde{g}{}_{5 4}
)(Y, U| X,|\boldsymbol{x}_T|,0 ;\tilde{T}_0 - \tilde{T}) 
%g{}_{\kappa 4}^{(0)}
%(Y, r, T| X,|{\boldsymbol x}_T|, T_0) 
%
\,d(\tilde{T}_0 - \tilde{T}),
%%%\kappa = (1, 4\hspace{0.2cm} \& \hspace{0.2cm} 5)
\end{split}
\end{equation}
that is now determined by (\ref{Hyp2}) when $\mathcal{F}(\tilde{\boldsymbol S}) = o(1)$ at arbitrary $\Omega=O(1)$ frequencies. Hence, setting the right hand side in (\ref{Hyp2}) equal to zero shows that the lowest order term in the expansion $\nu(y_1, r) = \bar{\nu}(y_1,r) + \bar{\nu}^{(1)}(y_1,r) +...$ is given by the solution to
\begin{equation}
\label{Hyp3}
\mathcal{L}
%{L}
%(y_1, U(y_1, r))
\bar{\nu}
(Y, U)
\equiv
\widetilde{c^2}
\frac{\partial }{\partial U}
\left(
\frac{1}{\widetilde{c^2}}
\bar{D}_0 \bar{\nu}
\right)
+
\bar{X}_1
\frac{\partial^2 \bar{\nu}}{\partial U^2}
=
0,
\hspace{0.25cm}
\textnormal{for}
\hspace{0.25cm}
\epsilon \ll O(1),
\end{equation}
where by the implicit function theorem, $\bar{\nu}(y_1,r)\equiv \bar{\nu}
(Y, U)\equiv  \widetilde{c^2} \bar{G}{}_{4} + \bar{G}{}_{5}
%(Y, U| X,|\boldsymbol{x}_T|,0 ; \Omega)
%\equiv
%\hat{\nu}{}^{ (0)} 
%(Y, U| X,|\boldsymbol{x}_T|,0 ; \Omega) 
%
%% 
$ is related to the zeroth-order azimuthal mode $\hat{\nu}{}^{ (0)} (Y, U)$ through the inverse Fourier transform of Eq. (\ref{nu_FT_azim}) in $(\Phi-\phi)$ where $(X,T_0) = \epsilon (x_1,t)$ are appropriate $O(1)$ slow variables for the observation field point $(x_1,t)$. %
Moreover, $Y = const.$ and ${dU}/{dY} = \bar{X}_1/U$ represent the characteristic curves (Garebedian 1998, pp. 121-122) of the hyperbolic second order PDE (\ref{Hyp3}). The pre-factor of the second member on the first line of (\ref{eq:Scaled_G}) allows the outer boundary, or matching, conditions (defined below) for the scaled inner solution $\bar{\nu} (Y, U)$ to depend on the observation point, $\boldsymbol{x}$, only through the polar angle, $\theta$.
%
% Note also that -iOmega coming from the FT of the far field solution to the wave equation (to which the adjoint equations reduce to) has been absorbed into solution \bar{nu}. It could have been written explicitly so as to cancel the same factor in the outer parallel flow solution. But we didn't do that.
%
The hyperbolic structure of (\ref{Hyp3}) shows that it is unnecessary to impose a downstream boundary condition.

Fig. 1 in GSA indicates how `$\bar{\nu}$-waves' propagate to both left and right from the $U = 0$ boundary and that no boundary conditions are required on the $Y = 0$ and $Y \rightarrow \infty$ boundaries (i.e. no inflow boundary condition is necessary). Thus, $\bar{\nu}(Y,U)$ is uniquely determined by the outer boundary conditions (by matching to the inner limit of the outer solution using the Van Dyke (1975) rule) obtained from the zero mean flow outer flow solution to (\ref{Hyp3}) when $\bar{X}_1 = 0$. That is, $   \bar{\nu}
%(Y,0) =
 \rightarrow
     -i\Omega c{}_\infty^2 e^{-i\Omega Y\cos\theta/c{}_\infty}$ and $   
     %{\partial\bar{\nu}}/{\partial U} 
     \bar{\nu}_U
     %(Y,0)= 
     \rightarrow
     -i\Omega c{}_\infty \cos\theta e^{-i\Omega Y\cos\theta/c{}_\infty}$
apply on the non-characteristic curve $U = 0$ where subscript denotes derivative and $U \rightarrow 0$ corresponds to the outer limit, $r \rightarrow \infty$. In these conditions, $Y\geq 0$ (note the sign error in Eqs. 5.45 $\&$ 5.48 in GSA)
and where $\theta$ is the polar observation angle from the jet centerline. 
Eq.(\ref{Hyp3}) and the matching condition above show that the composite Green's function $\bar{\nu}(Y,U; \Omega)$ is independent of jet spread-rate, $\epsilon$, at lowest order when the streamwise independent variable is taken to be $Y$ after the numerical solution to (\ref{Hyp3}) is determined in $(Y,U)$ co-ordinates  at fixed scaled frequencies, $\Omega$. 

For isothermal (or, slightly cold) jets, the temperature fluctuation $T^{\prime} \approx 0$, Afsar {\it et al}.(2019) show that $|e{}_{4 l}|/(c_\infty |e{}_{i l}|) \rightarrow 0$  (see line below \ref{eq:Rijkl}), so the $(\lambda = \mu=4) $component in the auto-covariance tensor ${R}_{\lambda j \mu l}
({\boldsymbol y}, {\boldsymbol \eta}; \tau)$ in (\ref{eq:Rijkl}) can be set equal to zero.
Hence reduction of the propagator from a $(4\times3)$ rank-2 tensor to a $(3\times3)$ one: ${\Gamma}_{\lambda, j}\rightarrow {\Gamma}_{i, j}$ in (\ref{eq:Iom2}) $\&$ (\ref{eq:Iom3}).
But the propagator (\ref{eq:Prop}) depends on $\bar{G}_\sigma (Y,r| {\boldsymbol x}; \Omega)$ and the mean flow (\ref{eq:Meanflow_exp}), therefore its solution must also separate out into the same asymptotic regions as in $\S$.2\ref{sec:2b} and depends on scaled variable/parameter combination $(Y, \Omega)=O(1)$.
The scaled propagator, defined in a similar manner to (\ref{eq:Scaled_G}), is then $\bar{\Gamma}_{\lambda, j}=\bar{\Gamma}_{\lambda, j}
(Y,r| {\boldsymbol x};\Omega)$. 
Taking the gradient operator, ${\boldsymbol\nabla} \equiv {\boldsymbol e}_1 {\partial}/ {\partial y_1}+ {\boldsymbol e}_r {\partial}/ {\partial r} + {\boldsymbol e}_\phi {\partial}/ {r \partial \phi}$  of the lowest order mean flow vector $\tilde{\boldsymbol{v}}({\boldsymbol y})$ in (\ref{eq:Meanflow_exp}) 
%$= U {\boldsymbol e}_1 + V_r {\boldsymbol e}_r$ 
we can easily show that the non-symmetric rank-two tensor, ${\partial \tilde{v}_\lambda}/{\partial y_j}$, in (\ref{eq:Prop}), where $\tilde{v}_\lambda \equiv \{ \tilde{v}_i, 0\}= \{ U, V_r, 0, 0\}$ possesses the following expansion: ${\partial \tilde{v}_i}/{\partial y_j} = (\partial U/\partial r) \delta_{i1} \delta_{jr} + O(\epsilon)$
in the $(Y,r, \phi)$ cylindrical co-ordinates using ${\partial {\boldsymbol e}_r}/ {\partial \phi} = {\boldsymbol e}_\phi$ $\&$ ${\partial {\boldsymbol e}_\phi} /{ \partial \phi} = -{\boldsymbol e}_r$.
Inserting this and the lowest order scaled Green's function vector, $\bar{G}_\sigma (Y,r| {\boldsymbol x}; \Omega) = \bar{G}_1 \delta_{\sigma 1}+\bar{G}_4 \delta_{\sigma 4}$ into (\ref{eq:Prop}) then shows that the latter possesses an asymptotic expansion,

\begin{equation}
%\begin{split}
\label{Prop_Exp}
\bar{\Gamma}_{i, j}
(Y,r| {\boldsymbol x};\Omega)
=
\delta_{i 1}
\delta_{j r}
\left(
\frac{\partial \bar{G}_1}{\partial r}
-
(\gamma-1)
\frac{\partial U}{\partial r}
\bar{G}_4
\right)
+
O(\epsilon)
,
%\end{split}
\end{equation}
in $(Y,r)$ co-ordinates at $\Omega=O(1)$ frequencies. 
$\bar{G}_\sigma (Y,r| {\boldsymbol x}; \Omega)$ is found in $(Y, U)$ co-ordinates using an equivalent re-scaling of the form (\ref{eq:Scaled_G}).
It is transformed back to $(Y=\epsilon y_1,r)$ co-ordinates for integration over $\boldsymbol{y}$ in (\ref{eq:Iom2}).
More specifically, since $\tilde{S}_i = \{\tilde{S}_1,\tilde{S}_r,\tilde{S}_5\}\equiv 0$ at lowest order, the solution to $\bar{\nu}(Y,U)$ via (\ref{Hyp3}) allows $\bar{G}_4$ to be determined using the $\tilde{D}_0 \tilde{\nu}$ relation defined above (\ref{hyb_eqn1}).
$\bar{G}_1$ is then determined using $\bar{G}_4$ in (\ref{G_1}) after replacing $\bar{G}_{5}$ with $\bar{G}_{4}$  and $\bar{\nu}$ (see also sentence below \ref{Hyp3}) 
where (\ref{G_1}) and $\tilde{D}_0 \tilde{\nu}$ are interpreted in terms of scaled Green's function variables using (\ref{eq:Scaled_G}) and $\partial \bar{G}_1/\partial r = (\partial U/\partial r)\partial \bar{G}_1/\partial U $ by the Chain rule.

\subsection{Low frequency acoustic spectrum formula}
\label{sec:acspecform}
\subsubsection{The standard approximations}
%\label{\sec:stand_approx}
It is well known (Leib $\&$ Goldstein 2011) that $\Gamma{}_{k, l}^*
({\boldsymbol y} + {\boldsymbol \eta}| {\boldsymbol x}; \omega)
$ can be approximated by taking advantage of the scale disparity  between the mean flow and turbulence relative to the acoustic wavelength $\lambda_a$ in the correlation volume $V({\boldsymbol \eta})$ of integral (\ref{eq:Iom2}) (but see Goldstein $\&$ Leib, 2018 for an analysis of the effects of azimuthal non-compactness).
In an asymptotic sense, the ALEE solution that enter $\Gamma{}_{k, l}^*$ via (\ref{eq:Prop}) will only contribute to integral over $O(|{\boldsymbol \eta}|)$ distances in (\ref{eq:Iom2}) 
when the mean flow length scales that determine the coefficients (and, therefore, the solution structure) of (\ref{eq:GAA})
%(i.e., a transverse mean flow length scale of $O(D_j)$ and a long streamwise scale of $O(1/\epsilon)$
%, where jet spread rate, $\epsilon \ll O(1)$) 
are of the same order as the turbulence correlation lengths in their respective directions.
This is because the latter propagator tensor, evaluated at $({\boldsymbol y} + {\boldsymbol \eta})$, multiplies $R_{i j k l}$ 
%%(including $\lambda=\mu=4$) 
in integral (\ref{eq:Iom2}).
%{\boldsymbol y} + {\boldsymbol \eta}
%
At minimum, the critical variation in $\Gamma{}_{\mu, l}^*$ occurs at the normalized far-field wavenumber $k_\infty\gg 1$, thus allowing $\Gamma{}_{\mu, l}^*$ to be represented by a Wentzel-Kramers-Brillioun-Jeffreys (WKBJ) approximation inasmuch as $\Gamma{}_{k, l}^*
({\boldsymbol y} + {\boldsymbol \eta} | {\boldsymbol x}; \omega) \approx\Gamma{}_{k, l}^*({\boldsymbol y} | {\boldsymbol x}; \omega) e^{i {\boldsymbol k}.{\boldsymbol\eta}}$.
Inserting this 
%the WKBJ approximation for $\Gamma{}_{\mu, l}^*$ 
into (\ref{eq:Iom3}) therefore gives an algebraic formula for the acoustic spectrum:
\begin{equation}
\label{eq:IomWKB}
I({\boldsymbol x}, {\boldsymbol y};\omega) 
\approx
(2\pi)^2
\Gamma_{i, j}
({\boldsymbol y}| {\boldsymbol x}; \omega)
\Gamma{}_{k, l}^*
({\boldsymbol y} | {\boldsymbol x}; \omega)
\Phi{}^*_{\lambda j \mu l}
({\boldsymbol y}, k_1, {\boldsymbol k}_T  ; \omega),
\end{equation}
where
\begin{equation}
\label{eq:Spec_Ten}
\Phi{}^*_{i j k l}
({\boldsymbol y}, k_1, {\boldsymbol k}_T ; \omega)
:=
\int
\limits_{V_\infty({\boldsymbol \eta})}
\mathcal{H}_{i j k l}
({\boldsymbol y}, {\boldsymbol \eta}; \omega)
e^{i {\boldsymbol k}.{\boldsymbol \eta} }
\,d{\boldsymbol \eta},
\end{equation}
such that the spectral tensor, $\Phi{}^*_{i j k l}$, possesses two-pair symmetries, $\Phi_{ij kl}=\Phi_{ji kl}=\Phi_{ij lk}$.

The final approximation we use is to allow the turbulence to be axisymmetric such that the transverse correlation lengths are small compared to that in the streamwise flow direction (Pokora $\&$ McGuirk measurements 2015, Fig. $19b$ cf. $20b$).
Afsar {\it et al}. (2011) used this data to propose that generalized auto-covariance tensor ${R}_{\lambda j \mu l}({\boldsymbol y},\eta_{_1}, \eta_{_\perp} ; \tau)$ is an axisymmetric tensor where  $\eta_{_\perp}=|{\boldsymbol \eta}_{_\perp}|$ and ${\boldsymbol \eta}_{_\perp}= (\eta_{_2}, \eta_{_3})$.
The spectral equivalent of this (lemma's $3.1$ and $3.2$ in Afsar 2012) requires that $\Phi{}^*_{\lambda j \mu l}({\boldsymbol y},k_{_1}, k{}_{_\perp}^2 ; \omega)$  is axisymmetric with the streamwise direction $k_{_1}$ being the principle direction of invariance. 
The physical space approximation is consistent with experiments by Morris $\&$ Zaman (2010) who show in Fig. 15 that the transverse and azimuthal correlation lengths are virtually constant across the Strouhal number range, $St=(0.01-1.0)$ for an isothermal axisymmetric jet.
Hence inserting (C.4) in Afsar {\it et al.} (2011) for the axisymmetric representation of $\Phi_{ij kl}$  then shows that the low frequency acoustic spectrum (\ref{eq:IomWKB}) can be approximated by one independent component of $\Phi_{\lambda j \mu l}$ as follow:
\begin{equation}
\label{I_low}
I({\boldsymbol x}, {\boldsymbol y};\omega) 
\rightarrow
\left(\frac{\epsilon}{ c{}_\infty^2 |\boldsymbol{x}|}\right)^2
%\left[
%
|\bar{G}_{12}|^2\Phi{}^*_{1212},
\end{equation}
where the tensor, $\bar{G}_{ij}$, is the symmetric part of the propagator tensor (\ref{Prop_Exp}). Suffix '$2$' in (\ref{I_low}) denotes the radial direction and its pre-factor is determined after inserting (\ref{Prop_Exp}) into the equivalent propagator version of the re-scaling of $\tilde{\nu}(Y,U)$ in Eq.(\ref{eq:Scaled_G}). Inasmuch as  $\tilde{\gamma}_{\lambda,j}(Y, U; T_0-T)$ appearing on the right hand side of (\ref{eq:Scaled_G}) and $\bar{\Gamma}_{\lambda,j}(Y, U)$ multiplied by this appropriate pre-factor on left side.
When the latter is inserted into (\ref{eq:IomWKB}), formula (\ref{I_low}) results.
The propagator in (\ref{I_low}) is therefore defined by an implicit function theorem-type statement (see paragraph above \ref{D0_trans}), which simply requires that
%shows that $G_{12}(y_1,r|{\boldsymbol x}; \omega)$ is 
%
\begin{equation}
\label{G_12}
G_{12}
(y_1,r,\psi|{\boldsymbol x}; \omega)
=
\tilde{G}_{12}
(Y(y_1), U(y_1,r))
=  
\frac{\partial \tilde{G}_1}{\partial r}
-
(\gamma-1)
\tilde{G}_4
\frac{\partial U}{\partial r}
\end{equation}
when $(\tilde{G}_1,\tilde{G}_4)$ and therefore $\tilde{\nu}(Y,U)$ are inserted into (\ref{eq:Scaled_G}).
It is worth noting that the numerical experiments conducted by G $\&$ L, Afsar (2010) on a parallel mean flow and Karabasov {\it et al}. (2010) on the full numerical solution to the ALEE (here written as \ref{eq:GAA}a - \ref{eq:GAA}c) corroborate the asymptotic
expansion (\ref{Prop_Exp}) and therefore (\ref{I_low}) in that the $G_{12}
(y_1,r,\psi|{\boldsymbol x}; \omega)$ propagator dominates the small angle acoustic radiation when inserted into (\ref{eq:IomWKB}).

\subsubsection{Experimentally verified model of ${R}_{1212}
({\boldsymbol y},\eta_1, |{\boldsymbol \eta}_\perp|, \tau)$ }
Since the linear transformation below (\ref{eq:Rijkl}) shows that, $\mathcal{H}_{1212} \equiv H_{1212}$, the spectral tensor component $\Phi{}^*_{1212}(\boldsymbol{y},k_1 ,k{}_T^2 ,\omega)$ is explicitly related to $R_{1212}$ via (\ref{eq:HFT}), (\ref{eq:Rijkl}) and space-time Fourier transform (\ref{eq:Spec_Ten}).
Our main focus in this paper is on the effect of non-parallelism on the propagator (\ref{eq:Prop}), we therefore use a previously successful model for  $\Phi{}^*_{1212}(\boldsymbol{y},k_1 ,k{}_T^2 ,\omega)$, which is a modification of Eq.(54) in Leib $\&$ Goldstein (2011).
Hence we allow ${R}_{1212}
({\boldsymbol y},\eta_1, \eta_T, \tau)$ to be represented by the following functional form (see also Afsar {\it et al}., 2017)
\begin{equation}
\label{eq:R1212_model}
%  \begin{split}
%
{R}_{1212}
({\boldsymbol y},\eta_1, |{\boldsymbol \eta}_\perp|, \tau)
%&
=
%\\
%&
{R}_{1212}
({\boldsymbol y},{\boldsymbol 0}, 0)
\left[
a_0
+ 
a_1
\tau
\frac{\partial}{\partial \tau}
+
a_2
\eta_1
\frac{\partial}{\partial \eta_1}
+
...
\right]
e^{-X(\eta_1, \eta_T, \tau)}
% \end{split}
\end{equation}
where the amplitude ${R}_{1212}
({\boldsymbol y},{\boldsymbol 0}, 0)$ 
is a function of ${\boldsymbol y}$ 
and is assumed to be proportional to the square of the local density weighted turbulence kinetic energy (see below).
%
%However, we do not include explicit convective streamwise variable $\eta_1 - U_c\tau$ (where $U_c$ is the convection velocity) in (\ref{eq:R1212_model}) or mixed higher order derivatives as Eqs. (47) $\&$ (48) in Leib $\&$ Goldstein.
%
%\ref{App:B} shows that (\ref{eq:R1212_model}) allows a more realistic estimation of $c_\perp$ compared to turbulence data. 
%
The leading term ($a_0$) in square brackets in (\ref{eq:R1212_model}) gives a  cusp for the auto-correlation of ${R}_{1212}
({\boldsymbol y},{\boldsymbol 0}, \tau)$ as $\tau \rightarrow 0$ and the derivative terms, (with coefficients, $a_1,a_2$), allows for anti (i.e. negative)-correlations with increasing $\tau$ and streamwise separation, $\eta_1$, respectively.
Leib $\&$ Goldstein (2011) show that the spectral function of the type $X(\eta_1, \eta_T, \tau) = \sqrt{{\eta{}_1^2}/{l{}_1^2} + {(\eta_1 - U_c\tau)^2}/{l{}_0^2} + f(\eta_T) }$ where $f(\tilde{\eta}_T)\sim \tilde{\eta}{}_T^4$ was found to best match Harper-Bourne's (2003) turbulence data ($\tilde{\eta}_i = {\eta_i}/l_i$, no sum on $l_i = (l_1, l_2, l_3)$).  
The length scales $l_i$ in this formula are taken to be proportional to the local turbulent kinetic energy, $k(\boldsymbol{y})$, and the rate of energy dissipation, $\tilde{\epsilon}(\boldsymbol y)$, determined via the RANS calculation; viz., $l_i = c_i (k^{3/2}/\tilde{\epsilon})(\boldsymbol y)$ 
where, suffix $i =(0,1,2,3)$, and, $c_i$, are now parameters that we can find by either comparing against experiment and/or LES data (see, for example, Fig.\ref{fig4_SPLpreds_comp}c).

Substituting (\ref{eq:R1212_model}) into (\ref{eq:Spec_Ten}) and performing an integration over $V_\infty (\boldsymbol \eta)$, Afsar {\it et al.} (2019) show, among other things, that the sound predictions when $\Phi{}^*_{1212}
({\boldsymbol y}, k_1, {\boldsymbol k}_T ; \omega)$ is inserted into (\ref{I_low}) are more-or-less identical to those obtained by allowing ${k}{}_T =0$ (with an error of $\ll 0.25$dB at $0.01<St<1.0$). We quote their final algebraic formula for $\Phi{}^*_{1212}$:
\begin{equation}
\label{eq:SpecPhi1212_A4}
% \begin{split}
%
%&
\frac{\Phi{}^*_{1212}
({\boldsymbol y}, k_1, 0 ; \omega)}{2\pi{R}_{1212}
({\boldsymbol y},{\boldsymbol 0}, 0)}
=
\frac{l_0 l_1 l{}_\perp^2}{\chi^2U_c}
%\\
%&
\left[
(1 - a_1 - a_2)
+
(a_1 \tilde{\omega}^2
-\bar{k}_1(\tilde{\omega}(a_1-a_2)l_1/l_0 -a_2\bar{k}_1))
\frac{4}{\chi}
\right]
% \end{split}
\end{equation}
where  we have put $ l_2 =l_3=l{}_\perp$ (which requires that $c_2 =c_3 = c_\perp$ in the length scale formula above) and $a_0=1$ so that ${R}_{1212}
({\boldsymbol y},0, 0, 0) /  {R}_{1212}
({\boldsymbol 0},\boldsymbol{0}, 0)= 1$. 
$\tilde{\omega} = \omega l_0/U_c$ is the non-dimensional frequency in (\ref{eq:SpecPhi1212_A4}) and $\chi(\bar{k}_1,\tilde{\omega})=
\bar{k}{}_1^2 + \tilde{\omega}^2 + 1 =
(\tilde{k}_1-(l_1/l_0)\tilde{\omega})^2 
+ \tilde{\omega}^2 + 1$.

Eq. (\ref{eq:SpecPhi1212_A4}) now depends on $6$ independent parameters: $(c_0, c_1)$; transverse length scale, $c_{\perp}$ and anti-correlation parameters: $(a_1, a_2)$ and the amplitude constant $a_{1 2 1 2}$ when we take $R_{1212} ({\boldsymbol y}, {\boldsymbol 0}; 0) = a_{1 2 1 2} \bar{\rho}^2 (\boldsymbol y) k^2(\boldsymbol y)$
where $\bar{\rho} (\boldsymbol y) k (\boldsymbol y)$ is the density-weighted RANS turbulent kinetic energy (TKE).
The amplitude pre-factor, $a_{1212}$, is usually approximated by its (maximum) value on the shear layer location $r = 0.5$ at the end of the potential core (Fig. 4 in Semiletov $\&$ Karabasov, 2016) but it could be measured in experiment if the cross-stream unsteady velocities can be measured or extracted via an LES calculation. 
Using the data in Fig. (4) of Karabasov {\it et al}. (2010), we take $a_{1212} = 0.25$ and $U_c = 0.68$ for all predictions in this paper. %Figs. (\ref{fig4_SPLpreds}) $\&$ (\ref{fig4_SPLpreds_comp}).
%1

\section{Analysis of (\ref{I_low}) and discussion of jet noise predictions}
\begin{figure}[H]
\begin{center}
%    \begin{subfigure} 
\includegraphics[width=0.35\textwidth]{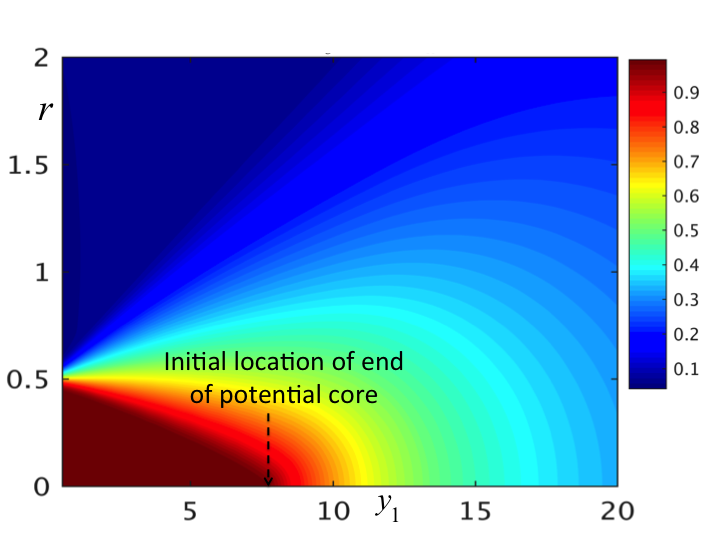}
        \label{fig_MF_1a}
%\end{subfigure}
%    \begin{subfigure}
\includegraphics[width=0.35\textwidth]{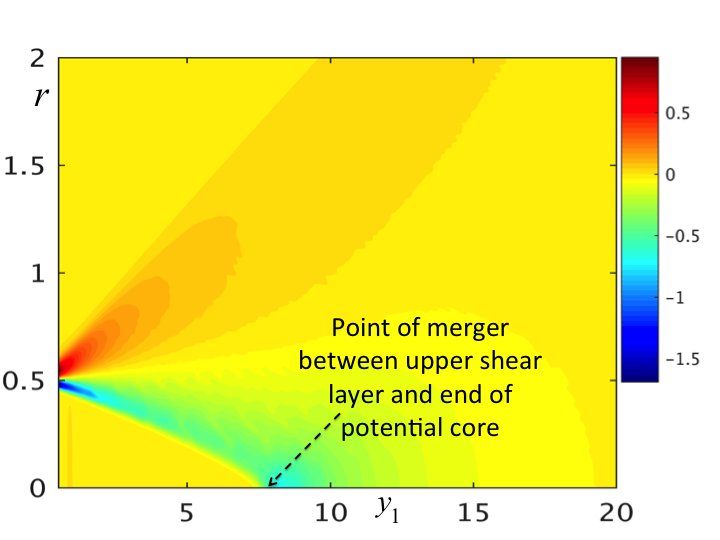} \\
\label{fig_MF_1b}
%\end{subfigure} \\
%(a). SP$07$: U($y_1$, r)  \hspace{80} (b). SP$07$: $V_r$($y_1$, r)  \hspace{60} (c). SP$07$: $\bar{X}_1$($y_1$, r)
(a). SP$07$: U($y_1$, r)  \hspace{30mm} (b). SP$07$: $\bar{X}_1$($y_1$, r)
\end{center}
\begin{center}
%    \begin{subfigure}
\includegraphics[width=0.35\textwidth]{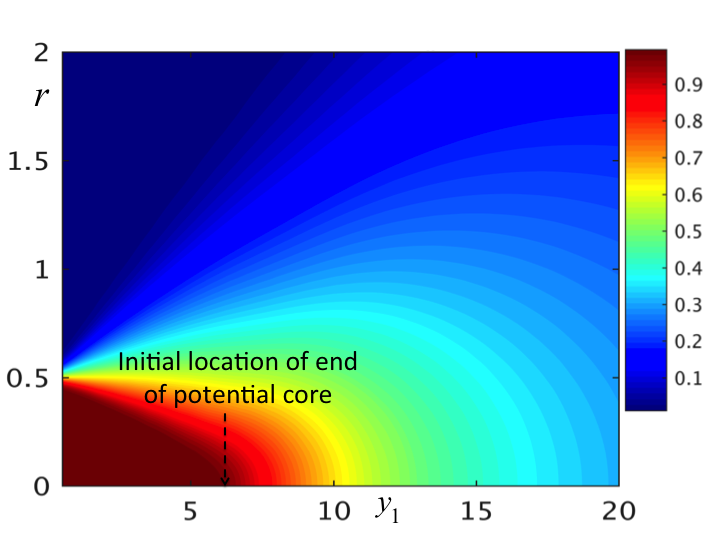}
\label{fig_MF_1c}
%\end{subfigure}
%    \begin{subfigure}
\includegraphics[width=0.35\textwidth]{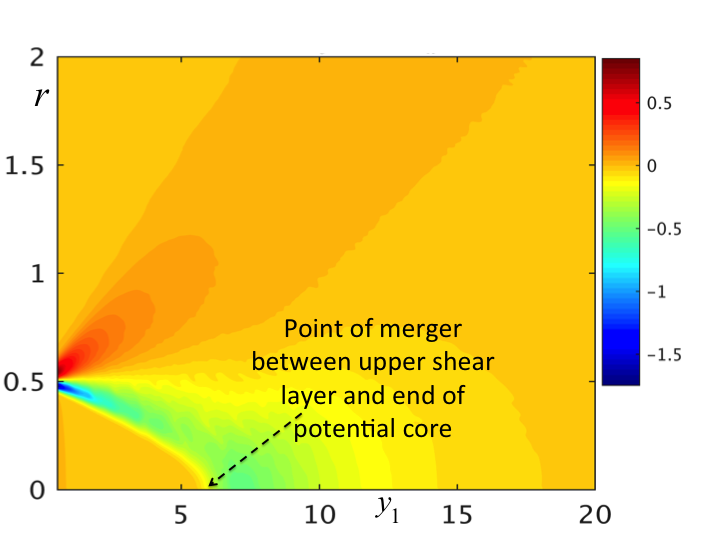} \\
\label{fig_MF_1d}
%\end{subfigure} \\
%(d)  \hspace{114} (e)  \hspace{114} (f)
(c). SP$03$: U($y_1$, r)  \hspace{30mm} (d). SP$03$: $\bar{X}_1$($y_1$, r)
%(d). SP$03$: U($y_1$, r)  \hspace{80} (e). SP$03$: $V_r$($y_1$, r)  \hspace{60} (f). SP$03$: $\bar{X}_1$($y_1$, r)
\end{center}
    \caption{Spatial distribution of mean flow components: $\tilde{v}_i=\{U, V_r\}$ and streamwise mean flow advection, $\bar{X}_1$ for SP$07$ ($Ma=0.9$ $\&$ $TR=0.84$) and SP$03$ ($Ma=0.5$ $\&$ $TR=0.95$).}
    \label{Fig1_meanflow}
\end{figure}

We analyze two axisymmetric jets in the Bridges (2006) data set at set points and flow conditions indicated in $\S.(1)$. 
The cases were chosen to highlight the effect of mean flow non-parallelism on the low frequency amplification of sound compared to predictions based on a locally parallel flow solution to the propagator $\bar{\Gamma}_{\lambda,j}(Y, U)$ in (\ref{I_low}).
For example, at the lower Mach number SP$03$ jet ($Ma =0.5$), the amplification in sound due to mean flow spreading is smaller than SP$07$ (see Fig. 5.2a in Afsar {\it et al}. 2016), which therefore results in a more broadband $30^o$ acoustic spectrum for SP$03$.
%prediction for the peak jet noise at $\theta = 30^o$, which is consistent with the data.
%
The mean flow field for the Green's function calculation in ${\Gamma}_{\lambda,j}(y_1, r)$ via (\ref{G_12}) is found from a steady RANS calculation using the {\sc Wind-US} code that was used in the Leib $\&$ Goldstein (2011) predictions. (The {\sc Wind-US} code was also validated against {\sc Fluent} solution in Afsar {\it et al}. (2019) for two supersonic acoustic Mach number jets in Bridges (2006)).
Comparing Figs.\ref{Fig1_meanflow}(a $\&$b) to \ref{Fig1_meanflow}(c $\&$ d) respectively shows that there is a reduction in jet potential core for SP$03$ compared to SP$07$ measuring about 20$\%$.
That is, as indicated in Fig.(\ref{Fig1_meanflow}), the initial normalized streamwise location $y_{_1}$ for the end of the potential core is $y_{_1}\approx7.5$ for SP$07$ and $\approx 6$ for SP03.
The greater spatial localization with reduced $Ma$ is also apparent in the spatial distributions of $\bar{X}_1$, which is the pre-factor that governs the effect of non-parallelism in the solution to $\bar{\nu}$ in (\ref{Hyp3}).
Indeed the streamwise location for the point of merger between the upper shear layer and the potential core in Figs.\ref{Fig1_meanflow}(b) $\&$ (d) (corresponding to the locus of maximum $|X_1|$) is reduced by more-or-less the same factor as the $U(y_1,r)$ contours in Figs.\ref{Fig1_meanflow}(a) $\&$ (c) are with their respective reduction in $Ma$.  

%\vspace*{-10.0pt}
%width was 0.32

\begin{figure}[H]
\begin{center}
% \begin{subfigure}
\includegraphics[width=0.4\textwidth]
{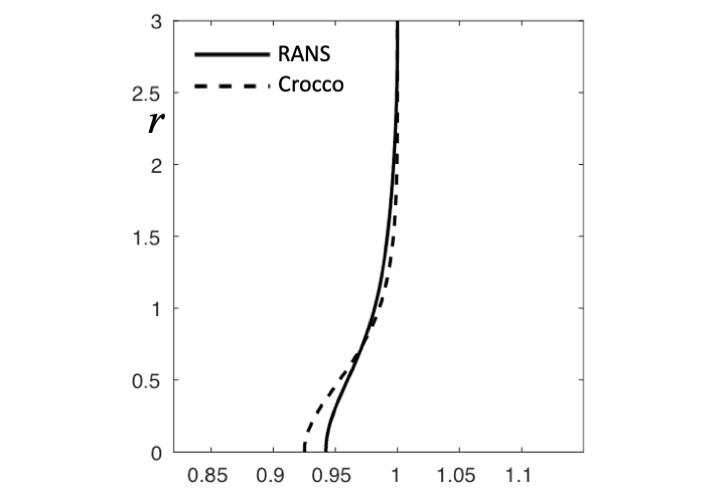} 
%{SP07p2_y1=10.png} 
\label{fig2a}
%\subcaption{(a).}
%\end{subfigure}
% \begin{subfigure}
\includegraphics[width=0.4\textwidth]{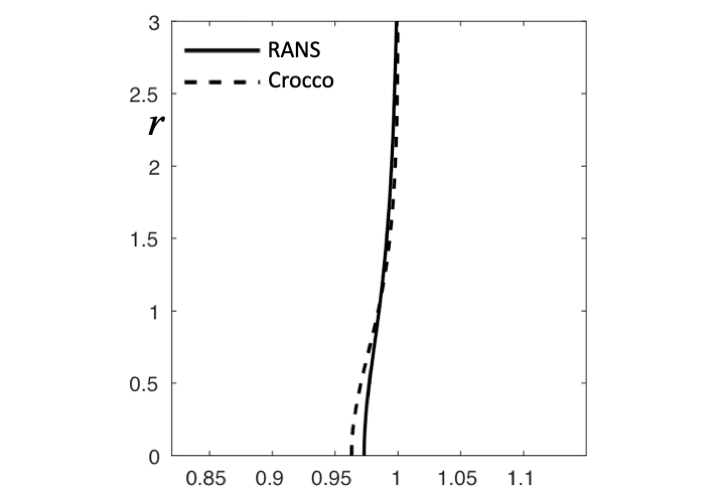} \\
\label{fig2b}
%\end{subfigure} \\
(a). SP$07$: $y_1 =10$  \hspace{40mm} (b).SP$03$: $y_1 =10$
\end{center}
 \caption{Verification of the Crocco relation $\widetilde{c^2}(y_1,r)/c{}_\infty^2$ (5.33 in GSA) against RANS mean flow for SP$07$ ($Ma=0.7$ $\&$ $TR=0.84$) and SP$03$ ($Ma=0.5$ $\&$ $TR=0.95$) respectively at $y_1 = 10$. Other points in the jet show similar trend.}
        \label{fig:Crocco+CB}
\end{figure}

%\vspace*{-10.0pt}
In Fig. (\ref{fig:Crocco+CB}) we compare the Crocco relation (defined below \ref{hyb_eqn1}) to the RANS-based $\widetilde{c^2}$ for SP$07$ and SP$03$.
%at $y_{_1} = 10$.
%%
The square brackets in the 
%$O(1)$ 
transformation 
%(i.e. prior to using method of multiple scales and matched asymptotic expansions) 
of the ALEE in (\ref{hyb_eqn1}) vanishes when $\widetilde{c^2}$  is defined in this way.    
The `worst' results are presented in Fig. (\ref{fig:Crocco+CB}). We show here that for both SP$07$ and SP$03$ there is maximum error of only 3$\%$ at the distant streamwise location of $y_1 = 10$ in Figs. (\ref{fig:Crocco+CB}a) $\&$ (\ref{fig:Crocco+CB}b) from the nozzle exit, which is already far downstream from the end of the potential core for both jets and especially so for SP$03$ (cf. Figs. (\ref{Fig1_meanflow}a) $\&$ (\ref{Fig1_meanflow}c) respectively).
At all other jet locations, i.e., $2\leq y_1 \leq14$, the difference in using Crocco's relation for $\widetilde{c^2}/c{}_\infty^2$ is less than $1 \%$ compared to that obtained from the RANS calculation.

%\vspace*{-5.0pt}

Since the components of the RANS mean velocity ($U$,$V_r$) are in a discrete form over a Cartesian mesh, the mapping between the $(Y,r)$ and $(Y,U)$ domains can no longer be done analytically as it was in GSA. 
Instead, the mapping is done numerically taking advantage of the monotone character of the section of the level curve $f(r) = U(Y=const., r)$ as $r\rightarrow\infty$ in the RANS solution. Thus, at any given $U$, a searching algorithm along a certain $Y=const.$ grid line is used to determine the corresponding  $r$ co-ordinate. 
Once this value of `$r$' is found, the derivatives $\partial U/\partial Y$ and $\partial U/\partial r$ are calculated by central differences in the original $(Y,r)$ grid; but the value of these derivatives are also the derivatives at the corresponding $(Y,U)$  point by implicit function theorem. 
Grid convergence investigations on the solution to $\bar{\nu}(Y,U)$ via  (\ref{Hyp3}) subject to matching conditions defined below this equation, are discussed in Afsar {\it et al}. (2019).
These results indicate that $\bar{\nu}(Y,U)$ remains reasonably converged when the above procedure is implemented numerically. 
For example, for a grid  of dimension $450\times300$ ($144,000$ points) there is only a very slight deviation appearing near the inner (jet) boundary, $U\rightarrow 1$ compared to one with 
$220,000$ points. Afsar {\it et al}. (2019) estimate this error to be less than 2$\%$. 

\begin{figure}[H]
\begin{center}
%     \begin{subfigure} 
\includegraphics[width=0.3\textwidth]{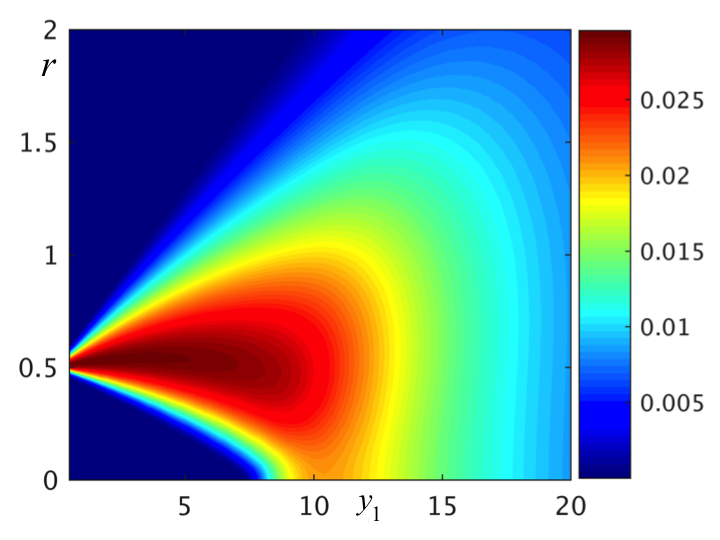}
        \label{fig3a}
%\end{subfigure}
%
%    \begin{subfigure} 
\includegraphics[width=0.3\textwidth]{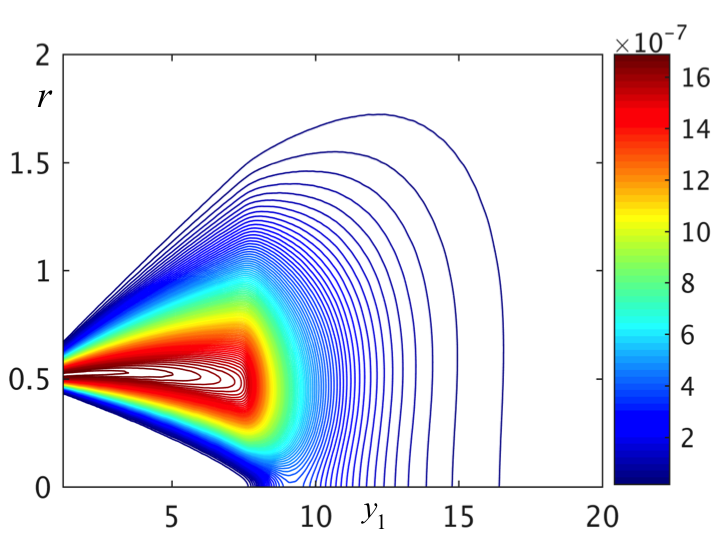}
        \label{fig3b}
%\end{subfigure}
%    \begin{subfigure}
\includegraphics[width=0.3\textwidth] {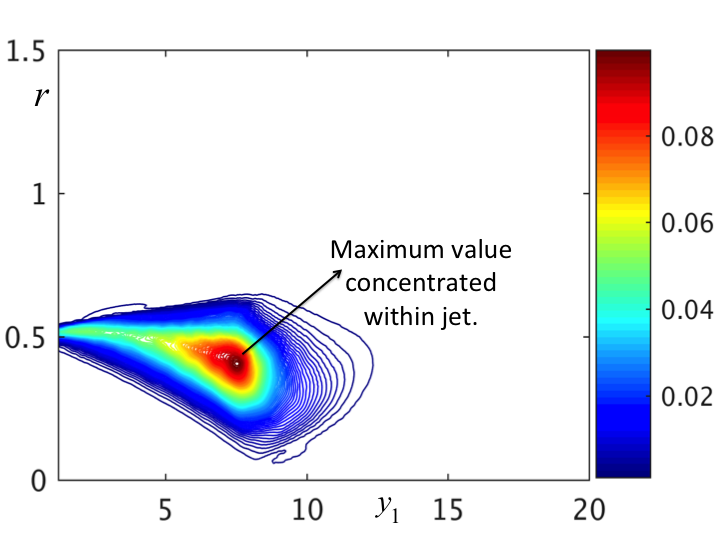} \\
\label{fig3c}
%\end{subfigure} \\
(a). SP$07$: k($y_1$, r)  \hspace{20mm} (b). SP$07$: $\Phi{}^*_{1212}$  \hspace{20mm} (c). SP$07$: $rI({\boldsymbol x}, {\boldsymbol y};\omega)$
\end{center}
\begin{center}
% \begin{subfigure} 
\includegraphics[width=0.3\textwidth]{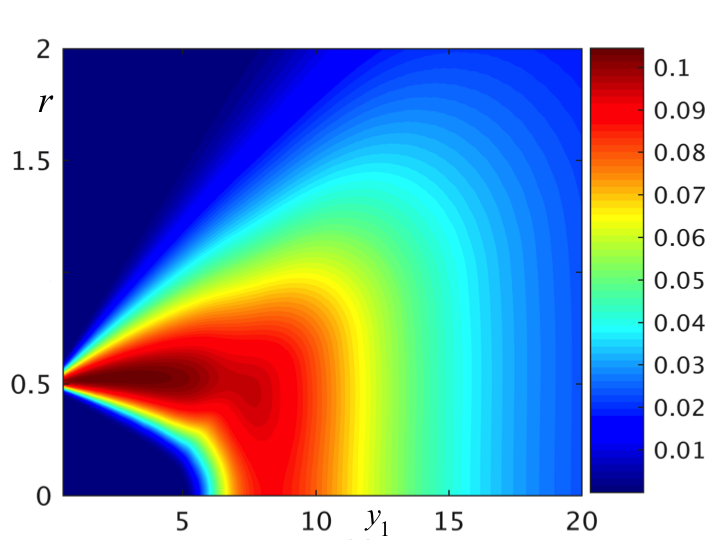}
        \label{fig3d}
%\end{subfigure}
%    \begin{subfigure}
\includegraphics[width=0.3\textwidth]{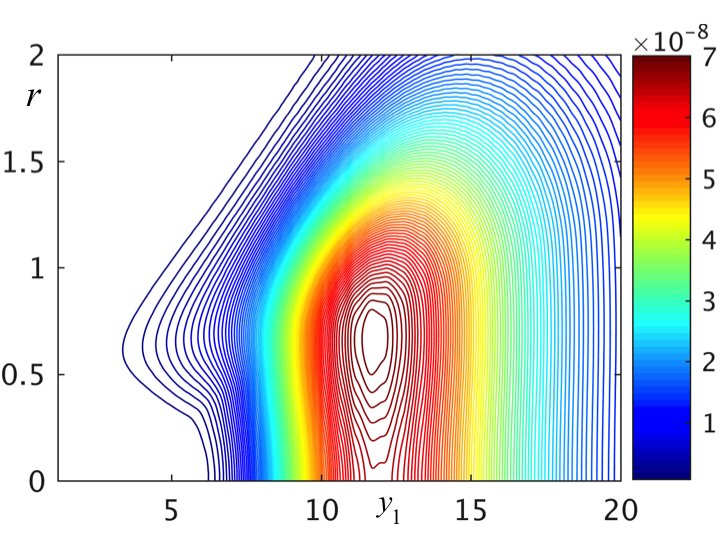}
\label{fig3e}
%\end{subfigure}
%    \begin{subfigure}
\includegraphics[width=0.3\textwidth]{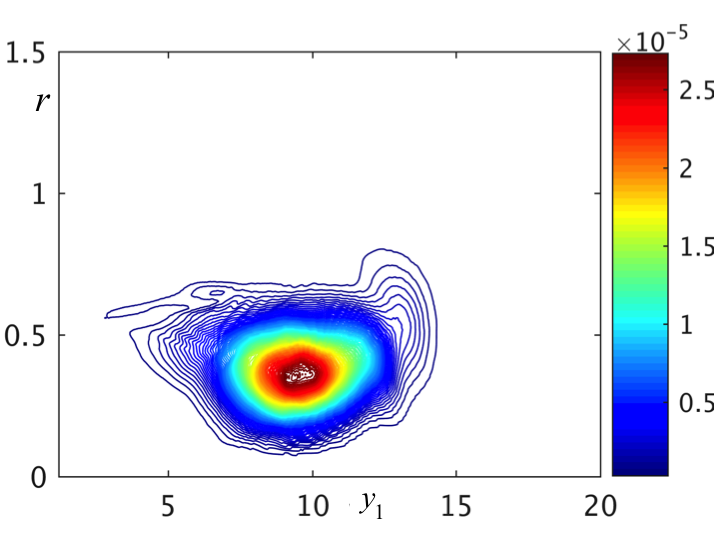} \\
\label{fig3f}
%\end{subfigure} \\
%(d)  \hspace{114} (e)  \hspace{114} (f)
(d). SP$03$: k($y_1$, r)  \hspace{20mm} (e). SP$03$: $\Phi{}^*_{1212}$  \hspace{20mm} (f). SP$03$: $rI({\boldsymbol x}, {\boldsymbol y};\omega)$
\end{center}
   \caption{Spatial distribution of Turbulent Kinetic Energy (TKE), $k(y_1,r)$, spectral tensor component $\Phi{}^*_{1212}
({\boldsymbol y}, k_1, k{}_T^2 ; \omega)$ given by (\ref{eq:SpecPhi1212_A4}) and $rI({\boldsymbol x}, {\boldsymbol y};\omega)$ via (\ref{I_low}) for SP$07$ and SP$03$ respectively. $St = 0.2$ and $\theta=30^o$. See Fig. (\ref{fig4_SPLpreds}) for turbulence scales used in (\ref{eq:SpecPhi1212_A4}). }
\label{Fig3:TKE}
\end{figure}

Figure (\ref{Fig3:TKE}) shows contours of $\Phi{}^*_{1212}
({\boldsymbol y}, k_1, 0 ; \omega)$  computed via (\ref{eq:SpecPhi1212_A4}). 
The radius-weighted acoustic spectrum, $rI({\boldsymbol x}, {\boldsymbol y};\omega)$, is found when this latter spectral model and the propagator component (\ref{G_12}) (determined via numerical solution to \ref{Hyp3}) is inserted into (\ref{I_low}).
%
%for SP$07$ and SP$03$respectively  
The frequency and far field location in Figs. \ref{Fig3:TKE} (b)-(c) $\&$ (e)-(f) correspond to the nominal peak noise location for SP$07$, namely $(St,\theta) = (0.2,30^o)$. 
The contours indicate that the turbulent kinetic energy $k(\boldsymbol y)$ and $\Phi{}^*_{1212}$ are an order of magnitude greater for SP$07$, which increases $rI({\boldsymbol x}, {\boldsymbol y};\omega)$ by almost $3$ orders of magnitude compared to SP$03$.
The large increase in $rI({\boldsymbol x}, {\boldsymbol y};\omega)$ for SP$07$ relative to SP$03$ can be explained by the amplification of $|\bar{G}_{12}|$ which at least for a parallel flow (inserting (7.2) in GSA into \ref{G_12}) is proportional to $|\partial U/\partial r|$ and is therefore more intense along the shear layer,  $r\sim0.5$, for SP$07$. 

\begin{figure}[H]
\begin{center}
%    \begin{subfigure} 
\includegraphics[width=0.35\textwidth]{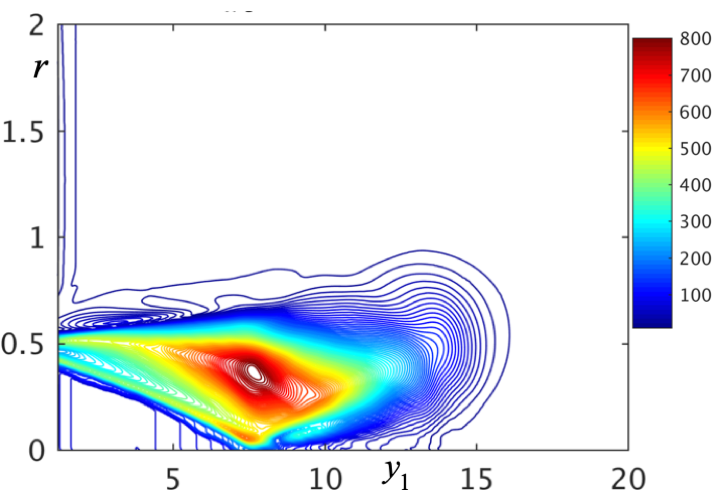}
        \label{fig4a}
%\end{subfigure}
%    \begin{subfigure}
\includegraphics[width=0.35\textwidth]{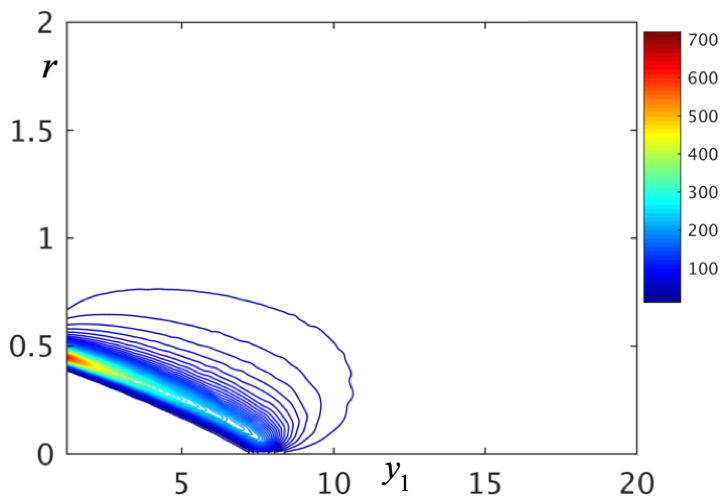} \\
\label{fig4b}
%\end{subfigure} \\
(a). SP$07$: (N-P)  \hspace{30mm} (b). SP$07$: (P)
\end{center}
\begin{center}
%    \begin{subfigure}
\includegraphics[width=0.35\textwidth]{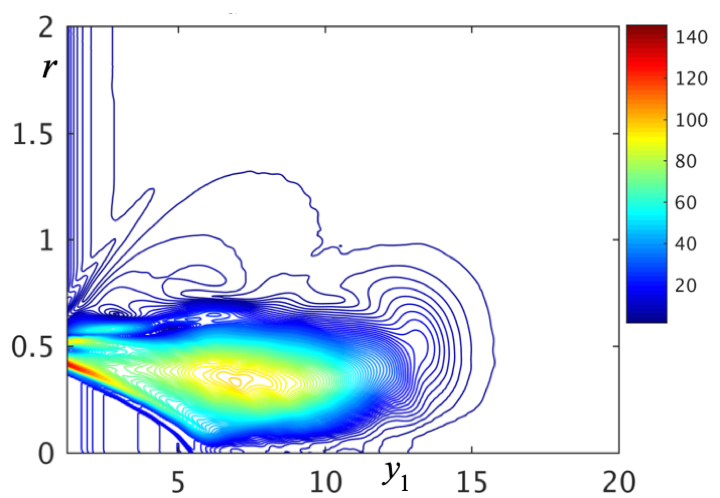}
\label{fig4c}
%\end{subfigure}
%    \begin{subfigure}
\includegraphics[width=0.35\textwidth]{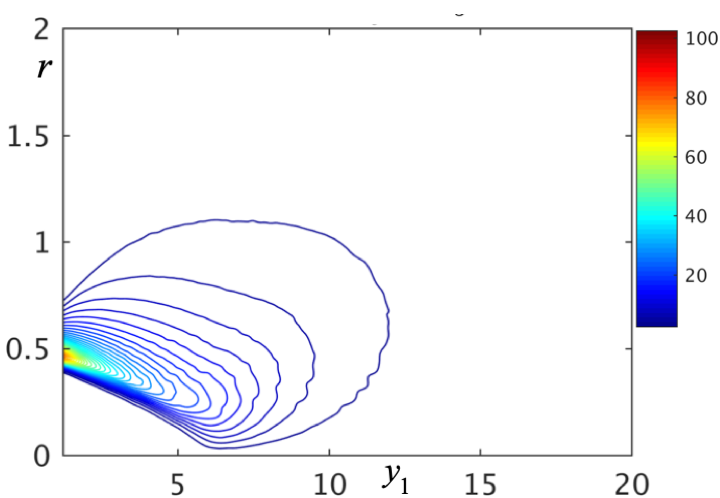} \\
\label{4d}
%\end{subfigure} \\
(c). SP$03$: (N-P)  \hspace{30mm} (d). SP$03$: (P)
\end{center}
    \caption{Spatial distribution of the momentum flux propagator $|d\bar{G}_1/dr|$ for SP$07$ ($Ma=0.9$ $\&$ $TR=0.84$) and SP$03$ ($Ma=0.5$ $\&$ $TR=0.95$) using the non-parallel (N-P) and parallel (P where $\bar{X}_1 =0$ in \ref{Hyp3}) flow solution to (\ref{Hyp3}).}
    \label{Fig4_dG1dr}
\end{figure}

%%%Here. Last bit
% figure labels AFTER captions
Spatial distribution of the momentum flux propagator $|d\bar{G}_1/dr|$ are shown in Fig.(\ref{Fig4_dG1dr}). This term forms the most intense part of $|\bar{G}_{12}|$ at small $\theta$. For SP$07$, $|d\bar{G}_1/dr|$ in non-parallel flow peaks at almost the same location as $rI({\boldsymbol x}, {\boldsymbol y};\omega)$ in Fig.(\ref{Fig3:TKE}c). 
The downstream peak in $|d\bar{G}_1/dr|$ is weaker for SP$03$ compared to SP$07$ (Figs.\ref{Fig4_dG1dr}a $\&$ c). It is interesting to note that the parallel flow computation of $|d\bar{G}_1/dr|$, essentially, has a single localized peak point at the nozzle lip for SP$07$; whereas for SP$03$, the contour lines $|d\bar{G}_1/dr|$ in parallel flow extend out across the outer edge of the jet shear layer (cf. in Fig.(\ref{Fig4_dG1dr}b) to Fig.\ref{Fig4_dG1dr}d). Since the spectral tensor component $\Phi{}_{1212}^*$ is large in this region of the jet (Fig.\ref{Fig3:TKE}e) such that the net effect of these extended contours in Fig.(\ref{Fig4_dG1dr}d) is to ensure that predicted sound using the parallel flow based Green's function in (\ref{I_low}) is similar to non-parallel at the low $Ma$ of SP$03$. 

%($Ma=0.9$ $\&$ $TR=0.9$) and SP$03$ ($Ma=0.5$ $\&$ $TR=1.0$) using the non-parallel (N-P) and parallel (P where $\bar{X}_1 =0$ in \ref{Hyp3}) flow solution to (\ref{Hyp3}).

%\vspace*{-20.0pt}
%\vspace*{-20.0pt}

\begin{figure}[H]
\begin{center}
%     \begin{subfigure} 
\includegraphics[width=0.32\textwidth]
{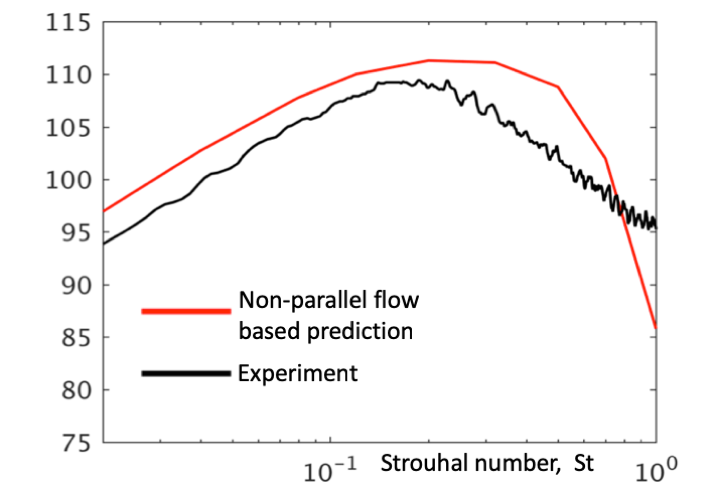}
        \label{fig5a}
%\end{subfigure}
%
%    \begin{subfigure} 
\includegraphics[width=0.32\textwidth]
{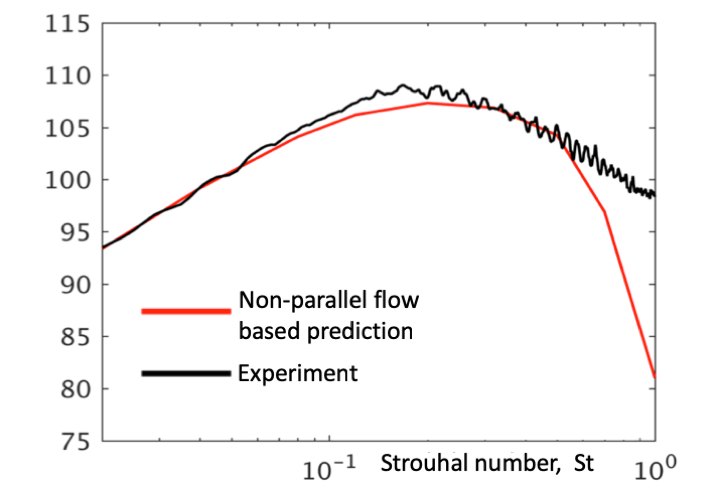}
        \label{fig5b}
%\end{subfigure}
%    \begin{subfigure}
\includegraphics[width=0.32\textwidth]{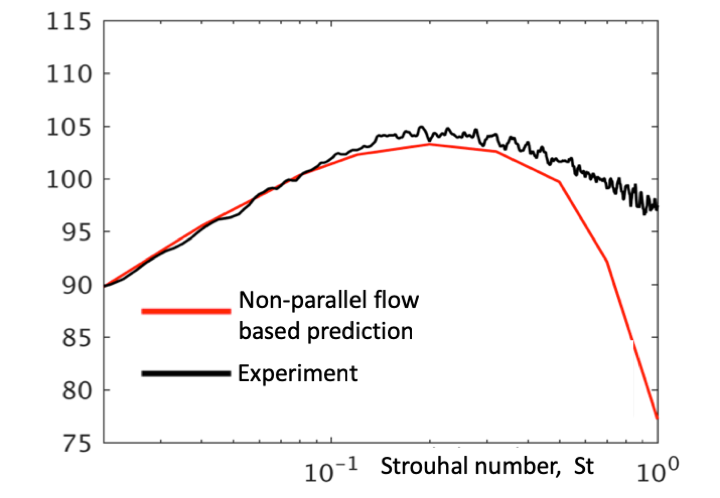} \\
\label{fig5c}
%\end{subfigure} \\
%(a)  \hspace{114} (b)  \hspace{114} (c)
(a). SP$07$: $\theta = 25^o$  \hspace{26mm} (b). SP$07$: $\theta = 30^o$  \hspace{26mm} (c). SP$07$: $\theta = 35^o$
\end{center}
\begin{center}
% \begin{subfigure} 
\includegraphics[width=0.32\textwidth]
{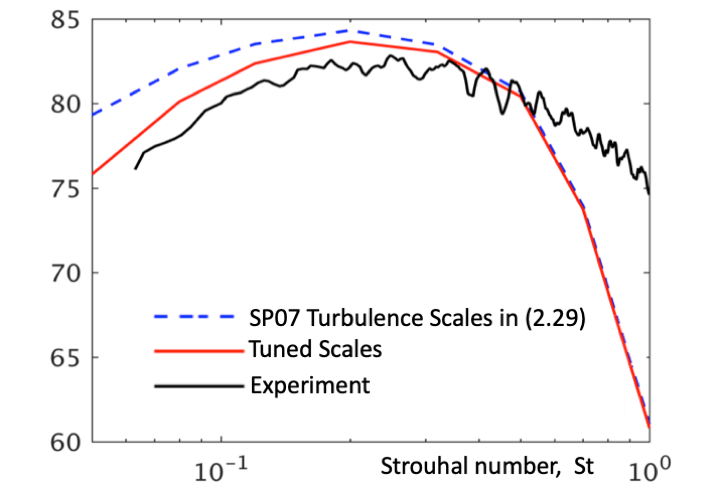}
        \label{fig5d}
%\end{subfigure}
%    \begin{subfigure}
\includegraphics[width=0.32\textwidth]
{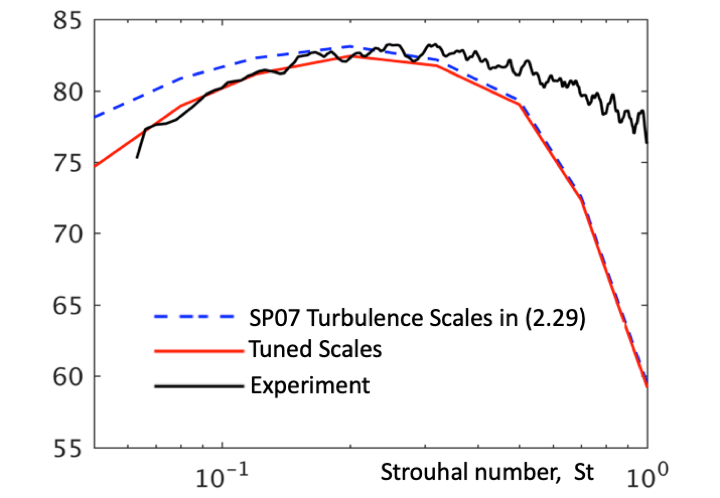}
\label{fig5e}
%\end{subfigure}
%    \begin{subfigure}
\includegraphics[width=0.32\textwidth]{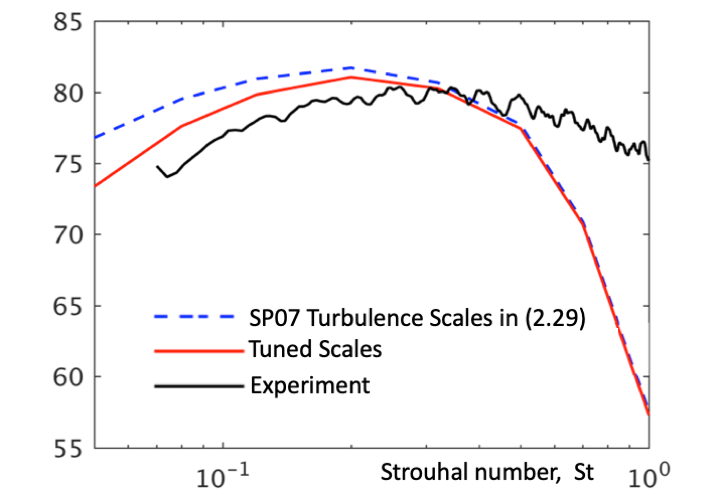} \\
\label{fig5f}
%\end{subfigure} \\
%(d)  \hspace{114} (e)  \hspace{114} (f)
(d). SP$03$: $\theta = 25^o$  \hspace{26mm} (e). SP$03$: $\theta = 30^o$  \hspace{26mm} (f). SP$03$: $\theta = 35^o$
\end{center}
   \caption{
   Power spectral density of acoustic pressure. Prediction compared with NASA experiments using acoustic spectrum formula (\ref{I_low}) in formula 
$SPL = 10log_{10} [4\pi(\rho U_J^2)^2 I({\boldsymbol x};\omega)/ P_{ref}^2]$ where $P_{ref}$ = $2\times 10^{-5}$ Pa. The spectrum $I({\boldsymbol x};\omega)$ is determined by integrating (\ref{I_low}) over \textbf{y} = ($y_1$, r, $\phi$) where turbulence scales in $\Phi_{1212}^*$ (\textbf{y}, $k_1$, $k_T^2$ ; $\omega)$ (\ref{eq:SpecPhi1212_A4}) are as follows ($a_1$, $a_2$; $c_0$, $c_1$, $c_{\perp}$): SP$07$ -  $(0.35,0.01; 0.2,0.15,0.02)$ ; Tuned predictions for SP$03$ - $(0.9,0.4; 0.2,0.15,0.016)$. Spread rate, $\epsilon = 0.09$ for both jets.}
\label{fig4_SPLpreds}
\end{figure}

The asymptotic scaling, $\omega=O(\epsilon)$, is expected to capture the dominant effects of non-parallelism 
%(via $\bar{X}_1$ in \ref{Hyp3}) 
in the lowest order propagator solution, $\bar{\Gamma}_{i, j}
(Y,r| {\boldsymbol x};\Omega)$ (\ref{Prop_Exp}), at small observation angles (typically for $\theta = 30^o$), 
%but also as $\theta\rightarrow 0$
we therefore consider predictions in the range of $\theta = (25^o, 30^o, 35^o)$ to assess the limit of its applicability.
Acoustic spectrum predictions are computed relative to the reference pressure after integrating (\ref{I_low}) over the source volume $dV_\infty( {\boldsymbol{y}})$ and inserting this into the formula in the caption to Fig.(\ref{fig4_SPLpreds}).
The results in Fig.(\ref{fig4_SPLpreds}) show that for the SP$07$ jet, predictions at the peak noise location of $\theta = 30^o$ remain quite accurate up to $St \approx 0.6$.
At an angle of $\theta = 25^o$ below this observation point, there is $\approx 5$ dB increase in the predicted power spectral density relative to the acoustic data (see Fig.\ref{fig4_SPLpreds}a).
While, away from the peak noise direction, at the angle $\theta = 35^o$, Fig.(\ref{fig4_SPLpreds}c) shows that predicted spectrum remains accurate within $(1-2)$ dB of the acoustic data upto the peak frequency (now at, $St \approx 0.4$) and thereafter rapidly decays.

\begin{figure}[H]
\begin{center}
% \begin{subfigure} 
\includegraphics[width=0.32\textwidth]
{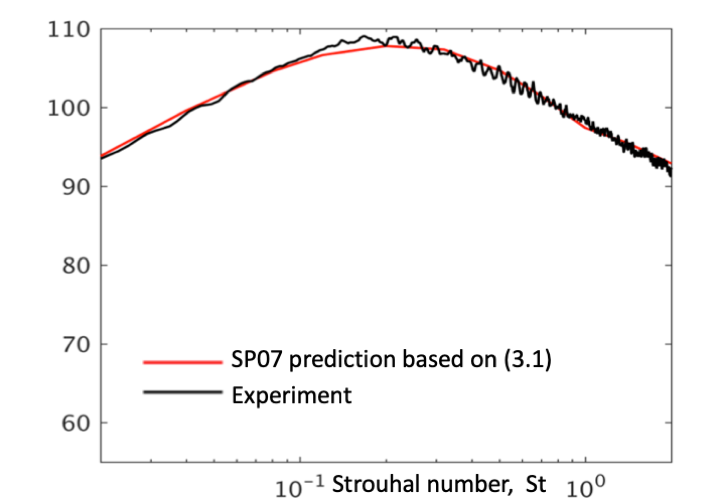}
%{SP07combined_30deg.png}
        \label{fig6a}
%\end{subfigure}
%    \begin{subfigure}
\includegraphics[width=0.32\textwidth]
{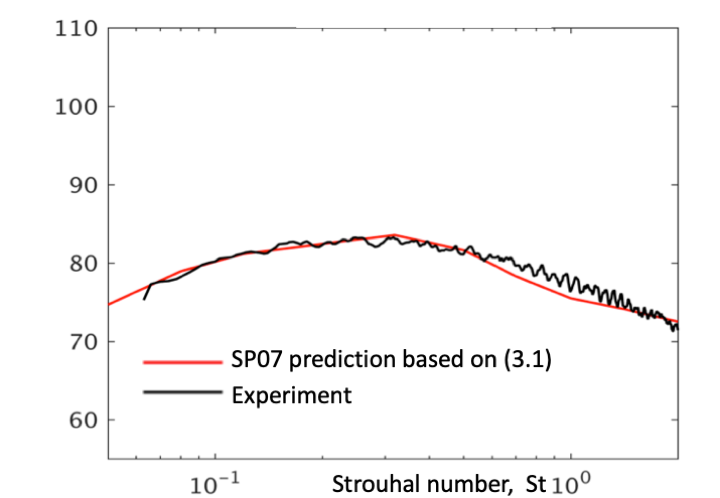}
%{SP03combined_30deg.png}
        \label{fig6b}
%\end{subfigure}
%    \begin{subfigure}
\includegraphics[width=0.32\textwidth]{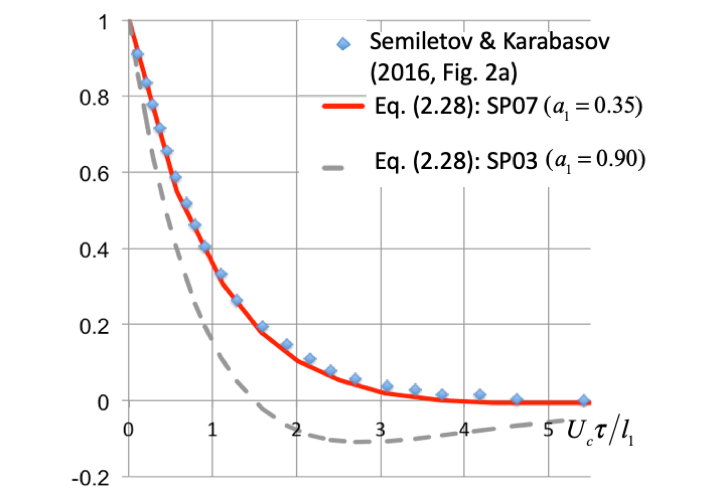} \\
%{Fig_5c.png}
        \label{fig6c}
%\end{subfigure} \\
%(a)  \hspace{114} (b)  \hspace{114} (c)
(a). SP$07$: $\theta = 30^o$  \hspace{26mm} (b). SP$03$: $\theta = 30^o$  \hspace{26mm} (c). ${R}_{1212}
({\boldsymbol y},0,0,  \tau)$
\end{center}
   \caption{
   Power spectral density (PSD) of acoustic pressure prediction compared with NASA experiments. PSD computed as caption as Fig. (\ref{fig4_SPLpreds}). Turbulence scales in $\Phi{}^*_{1212}
({\boldsymbol y}, k_1, k{}_T^2 ; \omega)$, (\ref{eq:SpecPhi1212_A4}) and (\ref{eq:nu_composite}) are as follows ($a_1$, $a_2$; $c_0$, $c_1$, $c_\perp$) : SP$07$ - (i). $St<St^{0}$, $(0.35,0.01; 0.2,0.15,0.02)$ ; (ii). $St>St^{0}$, $(0.35,0.01; 0.2,0.15,0.052)$. For SP$03$ - (i). $St<St^{0}$, $(0.9,0.4; 0.2,0.15,0.016)$ (ii). $St>St^{0}$,$(0.9,0.4; 0.2,0.15,0.04)$
Fig.(\ref{fig4_SPLpreds_comp})c shows validation of auto-correlation ${R}_{1212}
({\boldsymbol y},0,0,  \tau)$ against Semiletov $\&$ Karabasov (2016, Fig. 2a). 
}
\label{fig4_SPLpreds_comp}
\end{figure}

The parameters ($c_1/c_0, c_\perp,a_1, a_2$) were kept fixed for all cases and were chosen so that $(\eta_1-\tau)$ variation of  
${R}_{1212}
({\boldsymbol y},\eta_1, 0, \tau)$ agreed with LES data of the SP$07$ jet
at the end of the potential core (usually the region of maximum $k(y_1,r)$; see Fig.\ref{fig4_SPLpreds_comp}c).
Once $a_{1212}$ has been found in (\ref{eq:SpecPhi1212_A4}), via the LES data reported in Karabasov {\it et al}. (2010) for this case, the only element of empiricism, or hand-tuning, in our model is the estimation of $c_\perp$. 
(Note that parameters ($c_1/c_0, a_1, a_2$) were found by comparison of (\ref{eq:R1212_model}) to Fig. $2a$ in Semiletov $\&$ Karabasov (2016). This requires $c_1$ or $c_0$ to be set once the ratio $c_1/c_0$ is fixed).
The latter parameter, $c_\perp$, enters (\ref{eq:SpecPhi1212_A4}) as pre-factor (since $l_\perp \propto c_\perp$) and therefore governs the absolute level of the acoustic predictions by affecting the amplitude of $\Phi{}^*_{1212}
({\boldsymbol y}, k_1, 0 ; \omega)$.
We found that a value of $c_\perp$ (see caption of Fig. \ref{fig4_SPLpreds}), more-or-less an order of magnitude smaller than the streamwise turbulence length scale parameter $c_1$, gave best agreement for the $\theta = (25^o, 30^o, 35^o)$ cases.
Relative values of the transverse and streamwise correlation lengths such as this are consistent with the turbulence measurements of Morris $\&$ Zaman (2010, Table $4$) and Pokora $\&$ McGuirk (2015, Fig. 19b cf Figs. (20-21)b) which also agree with higher $Ma$ LES data in Karabasov {\it et al}. (2010, see their Fig. 6).  
Further evidence for this scale reduction in the transverse direction of a higher-order correlation function was given by Harper-Bourne (2003). Comparing Figs.$(7b)$ $\&$ $(8b)$ in his paper shows that the ratio of correlation lengths in the streamwise $(L_1)$ to transverse directions $(L_2)$ was $L_1/L_2 \approx 7.3$ in the correlation of $v{}_1^{\prime 2}$ for the low Mach number axisymmetric jet he considered.
While the correlation of $v{}_1^{\prime 2}$ (the equivalent of $R_{1111}$ in our notation) is not the same as $R_{1212}$, it is expected to have similar space-time decay when  appropriately normalized, as Semiletov $\&$ Karabasov's (2016) work showed. Hence we can expect Harper-Bourne`s results for the scale reduction in the transverse direction to give at least a ball-park figure for $c_1/c_\perp$. 
The ratio of $c_1$ to $c_\perp$ that we used in the predictions for SP$07$ in \ref{fig4_SPLpreds}(a-c) is quite similar, at $c_1/c_\perp = 7.5$.

As $\theta$ increases the overall level of the predictions lie below the acoustic data for higher frequencies at $\theta = 35^o $ for example. This behavior continues for larger angles $45^o$ (not shown here) but even in these higher $\theta$  cases the spectral shape of the prediction up to the peak frequency is more-or-less parallel to the acoustic data.
In other words, the low frequency roll off is well predicted albeit positioned lower than the acoustic data.
Hence an increase in $c_\perp$ to bring the predictions in line with the data would achieve very good accuracy even upto $\theta = 45^o$ also. But this  would be at the expense of physical consistency of the model for
${R}_{1212}
({\boldsymbol y},\eta_1, |{\boldsymbol \eta}_\perp|, \tau)$,  since it would require that $c_\perp \sim c_1$ which is contrary to the structure of axisymmetric tubulence observed in experiments and LES (Morris $\&$ Zaman, 2010; Karabasov {\it et al}., 2010; Pokora $\&$ McGuirk, 2015) discussed above.

The property that the low frequency noise amplification due to non-parallel flow effects is dominant (or, $O(1)$) at small $\theta$ only was found by Karabasov {\it et al}. (2010) who showed that at $\theta \geq 60^o$ acoustic predictions using the full numerical solution of the ALEE (\ref{eq:GAA}) is basically identical to that obtained when locally parallel mean flow approximation, $\tilde{v}_i = \delta_{i1} U(r)$, of the ALEE are solved and inserted into the propagator (\ref{eq:Prop}) (see their Fig. 16).
Our $\bar{\nu}(Y,u)$ solution naturally recovers Karabasov's result as $\theta \rightarrow 90^o$ since $\bar{\nu}(Y,U)\rightarrow \bar{\nu}^{p}(Y,U) = const.$ at this angle and inasmuch as $|\bar{\nu}^{p}(Y,U;\theta = 90^o)|  \ll |\bar{\nu}(Y,U;\theta = 30^o)|$  where $\bar{\nu}^{p}(Y,U)$ is the locally parallel solution to (\ref{Hyp3}) and is given by (7.1) in GSA. 
It is easy to prove this because a non-parallel flow solution of the form $\bar{\nu}^{\dag}
(Y, U)$ will always be zero when the decomposition $\bar{\nu} = \bar{\nu}^{p} + \bar{\nu}^{\dag}$ is inserted into (\ref{Hyp3}) and the limit $\theta\rightarrow 90^o$ is taken.
The locally parallel flow solution $\bar{\nu}^{p}$ satisfies $\partial( \bar{D}_0 \bar{\nu}^{p}/\widetilde{c^2})/\partial U = 0$ by definition; then $\bar{\nu}^{\dag} \rightarrow 0$ since the latter, determined by (\ref{Hyp3}), is subject to the reduced  outer boundary conditions  $\bar{\nu}^{\dag}(Y,0)\rightarrow const.$ and $\bar{\nu}{}_{U}^{\dag}(Y,0)\rightarrow 0$ as $\theta\rightarrow 90^o$. 
In other words the only compatible inner solution that can match onto the outer parallel flow solution as $U\rightarrow 0$ and $\theta \rightarrow 90^o$ is a constant. Our numerical simulations confirm this.
%
%\vspace*{-20.0pt}

%\vspace*{-10.0pt}

%
We, initially, kept the turbulence scales ($c_1/c_0, c_\perp,a_1, a_2$) in (\ref{eq:SpecPhi1212_A4}) fixed at the same value as that for SP$07$ for the SP$03$ predictions (see caption of Fig. \ref{fig4_SPLpreds}).
That, however, resulted in a $(3-5)$dB over-prediction of the $30^o$ spectrum in the low frequency region, $0.01<St<0.15$ where we expected the theory to be accurate. 
%for which the non-parallel flow theory to agree
This was remedied by an increase in the anti-correlation parameters $(a_1, a_2)$ to $(0.9, 0.4)$ in (\ref{eq:R1212_model}) $\&$ (\ref{eq:SpecPhi1212_A4}) from a value of $(a_1, a_2) = (0.35, 0.01)$. To re-iterate, the latter set of scales were used for SP$07$ and agreed with Semiletov $\&$ Karabasov turbulence data for the same jet (see our Fig. \ref{fig4_SPLpreds_comp}c).
In Figs. (\ref{fig4_SPLpreds}) (d-f) we show the SP$03$ predictions at $\theta = (25^o, 30^o, 35^o)$ using SP$07$ turbulence scales in addition to those obtained by via $(a_1, a_2) = (0.9,0.4)$. The $(c_0,c_1)$ parameters in (\ref{eq:SpecPhi1212_A4}) for the SP$03$ jet were nonetheless kept fixed to what we found for SP$07$.
While there is a region of agreement at $St\leq 0.3$ and $\theta = 30^o$ between our prediction and the acoustic data  for SP$03$ when  $(a_1, a_2) = (0.9,0.4)$, this results in an anti-correlation region of amplitude $\sim0.1$ in $R_{1212}$ (see Fig. \ref{fig4_SPLpreds_comp}c).
We note here also, that in the results shown in Fig.(\ref{fig4_SPLpreds}), the calculations were run at a fixed jet spread rate, $\epsilon=0.09$, although in reality the lower-$Ma$ SP$03$ jet will spread at a faster rate than the SP$07$ one (cf. Figs.\ref{Fig1_meanflow}a $\&$ \ref{Fig1_meanflow}c), perhaps contributing to the need for modified source parameters to obtain a better fit in the former case. 

There are two additional possible explanations for why the predictions in the SP$03$ case are not as close to the data as they are for SP$07$ when the same turbulence parameters are used to model ${R}_{1212}
({\boldsymbol y},\eta_1, |{\boldsymbol \eta}_\perp|, \tau)$ in (\ref{eq:R1212_model}). 
First, it could be that turbulence structure for SP$03$ is dissimilar to SP$07$.
But Semiletov $\&$ Karabasov's (2016) LES calculation of ${R}_{1111}
({\boldsymbol y},\eta_1, 0, \tau)$ agreed with Harper-Bourne's low Mach number ($Ma = 0.22$) turbulence measurements of the same component (this was also confirmed in Karabasov {\it et al}. 2010 and Pokora $\&$ McGuirk, 2015, Fig. $19b$).
Hence without numerical and/or experimental confirmation, it would seem reasonable to suggest that ${R}_{1212}
({\boldsymbol y},\eta_1, 0, \tau)$ does not possess an anti-correlation (negative) region -- however small -- in its auto-correlation for SP$03$. 
%is not consistent with how we expect the turbulence structure of this jet to look like.
%
The only alternative explanation is that there is no non-uniformity in the $\tilde{\nu}-$solution at low $Ma$. This means that ${\boldsymbol g}{}_{4}^{a} ({\boldsymbol y}, \tau| {\boldsymbol x}, t)$ depends on $(t-\tau)=O(1)$ and not the slow time $(\tilde{T}_0 - \tilde{T}) = O(1)$ when $|{\boldsymbol x}|$ is in the peak noise location.
But since the non-parallel flow theory does not reduce uniformly to the locally parallel flow solution when $Ma\ll 1$, the direct parallel flow solution to (\ref{eq:GAA}) would therefore be more appropriate than solving (\ref{Hyp3}) for the Green's function of the SP$03$ jet (see also Fig.\ref{Fig4_dG1dr} and associated discussion).
This is because non-parallelism does not diminish fast enough in (\ref{Hyp3}) when $Ma\ll 1$ owing to the residual effect of $\bar{X}_1$ that remains at $\bar{X}_1 = O(Ma)$ in the inner region and which therefore alters the solution, $\bar{\nu}(Y,U)$, to (\ref{Hyp3}).
Notwithstanding the fact that the direct solution of ALEEs at $Ma =0.5$ of SP$03$ has not been performed showing whether any amplification in the acoustic spectrum exists at low/small $(St,\theta)$ or not when a non-parallel flow Green's function is used in (\ref{I_low}) compared to that obtained by the locally parallel flow, this latter effect is probably not justified when $Ma\ll 1$.
It is also worth noting that at the outer boundary $\bar{X}_1(Y,U)$ is asymptotically small inasmuch as $\bar{X}_1= o(U)$ as $U\rightarrow 0$ no matter what the acoustic Mach number is (meaning, $Ma=O(1)$); this therefore implies that, $\bar{\nu}= V_0(Y) + o(U)$ where, $V_0(Y)$, is identical to the inner limit of the parallel flow solution to (\ref{Hyp3}) found by letting $U\rightarrow 0$ in $(7.2)$ of GSA (see also p.$13$ of their paper).

In Fig. (\ref{fig4_SPLpreds_comp}) we, therefore, show predictions based on an approximate composite formula for $\bar{\nu}(Y,U)$ in which the non-parallel flow based Green's function is used below the peak frequency and the sum of the parallel and non-parallel used at higher frequencies after the peak. 
In other words in $(y_1,r;\omega)$ co-ordinates:
\begin{equation}
\label{eq:nu_composite}
\bar{\nu}(y_1,r;\omega) \rightarrow \bar{\nu}(y_1,r;\omega) + H(\omega - \omega^{(0)})\bar{\nu}^{p}(y_1,r;\omega)
\end{equation}
where $\bar{\nu}^{p}(y_1,r;\omega)$ is the locally parallel flow solution to (\ref{Hyp3}), $\bar{\nu}(y_1,r;\omega)$ is determined by solution to (\ref{Hyp3}) and its matching conditions; $H(\bullet)$ is the Heaviside function of stated arguments and $\omega^{(0)}$, the peak frequency.
The acoustic predictions determined using the approximate composite Green's function formula (\ref{eq:nu_composite}) gives excellent agreement for the $30^o$ spectrum at all frequencies $0.01\leq St \leq 2.0$ for both SP$07$ and SP$03$. 
The turbulence parameters used for these predictions are given in the caption to Fig.(\ref{fig4_SPLpreds_comp}). 
For SP$07$, the turbulence scales ($c_1/c_0, c_\perp,a_1, a_2$) for $St <St^{(0)}$ (peak Strouhal number) were kept the same as that used in Fig. (\ref{fig4_SPLpreds})(a-c) which, as stated, were found to be consistent with LES-determined turbulence simulations in Fig.(\ref{fig4_SPLpreds_comp}c).
At high frequencies, $St>St^{(0)}$, we found that an increase in $c_\perp$ was necessary for SP$07$ from $c_\perp = 0.02$ to $c_\perp = 0.052$ for the composite Green's function that now involves the sum,  
$\bar{\nu}(y_1,r;\omega) + \bar{\nu}^{p}(y_1,r;\Omega)$ where the non-parallel flow term, $\bar{\nu}(y_1,r;\omega)$, exponentially decays. 
But this still satisfies the consistency requirement that $c_\perp \ll c_1$) (where $c_1/c_\perp = 2.9$ in this case).
The predictions for SP$03$ at $St>St^{(0)}$ in Fig.\ref{fig4_SPLpreds_comp}b) also required an increase in $c_\perp$ ($c_\perp=0.04$) compared to what we used in Fig. (\ref{fig4_SPLpreds})(d-f) (where $c_\perp = 0.016$). Here, too, $c_1/c_\perp >1 = 3.8$. 
%the latter of which, as we have indicated, introduces a small anti-correlation region into the model for ${R}_{1212}$ owing to values $(a_1, a_2)$ that were needed to obtain good agreement the acoustic data.
%

%
%\vspace*{-4pt}

\vspace*{-20pt}
\section{Conclusion}

Non-parallel flow effects will enter the lowest order expansion of the adjoint linearized Euler equations (ALEE), (\ref{eq:GAA}), when the temporal evolution of the adjoint vector Green's function is slow and of the same order as the asymptotically small jet spread rate.
Goldstein, Sescu $\&$ Afsar (GSA, 2012, Fig. 25) showed that this distinguished limit introduces qualitatively similar structure as the full numerical solution to the ALEE computed by Karabasov {\it et al.} (2010, 2013) at the peak noise location.
Afsar {\it et al}. (2019) showed that the GSA theory can be more easily derived by taking the streamwise mean flow component, $U(y_1,r)$ as one of independent variables of choice in the ALEE, (\ref{eq:GAA}), prior to any asymptotic analysis.
This results in mixed partial differential equation, (\ref{Hyp2}), for the adjoint vector Green's function when the Favre-averaged speed of sound $\widetilde{c^2}$ is assumed to satisfy Crocco's relation (which our subsequent numerical checks in Fig. (\ref{fig:Crocco+CB}) support).
%B
But the right hand side of (\ref{Hyp2}) drops out of lowest order expansion  when the mean flow field is slowly varying (\ref{eq:Meanflow_exp} $\&$ \ref{eq:X_exp}) with small spreadrate, $\epsilon$. 
By dominant balance considerations, the appropriate asymptotic expansion of the radial and azimuthal components of the vector Green's function of the momentum equation (Eq. $2.13a$ in Goldstein, 2003) ensures that the final hyperbolic equation (\ref{Hyp3}) is identical  Eq. $(5.31)$ in GSA (see Afsar {\it et al.} 2019).
The dependent variable that this hyperbolic operator acts upon is the combined Green's function $\bar{\nu} = \widetilde{c^2} \bar{G}_4 + \bar{G}_5$, where by (\ref{eq:GFT}) and (\ref{eq:Scaled_G}), ($\bar{G}_4$, $\bar{G}_5$) correspond to scaled Fourier transforms of the adjoint Green's functions for the energy and continuity equations (respectively given by Eqs. 2.13a $\&$ 2.9a in Goldstein, 2003).
%\ref{eq:GAA} b $\&$ c 
%
The solution, $\bar{\nu}$, is then subject to appropriate matching conditions on the non-characteristic curve, positioned in the outer region at $U=0$ and defined below (\ref{Hyp3}).

Equation (\ref{Prop_Exp}) shows that introducing the `synchronized' low frequency/small spread rate asymptotic scaling into the propagator tensor of the generalized acoustic analogy (\ref{eq:Prop}) gives a prediction formula for the acoustic spectrum that depends on a single spectral tensor term $\Phi{}^*_{1212}$ in (\ref{I_low}) when consistent approximations are made to model the turbulence auto-covariance tensor (\ref{eq:Rijkl}). %
$\Phi{}^*_{1212}(\boldsymbol{y},k_1 ,k{}_T^2 ,\omega)$ is related to the real space tensor component, ${R}_{1212}
({\boldsymbol y},\eta_1, |{\boldsymbol \eta}_\perp|, \tau)$ via the $4-$dimensional space-time transform (\ref{eq:HFT}) $\&$ (\ref{eq:Spec_Ten}).
Fig. (\ref{fig4_SPLpreds}) reveals that acoustic predictions based on this asymptotic approach remain within $(1-2)$ dB of the acoustic data for the high speed SP$07$ ($Ma = 0.9$) jet at the polar observation angles, $\theta = (30^0, 35^0)$ that correspond to the range of observer locations where the peak sound is measured.
The turbulence model we constructed for the ${R}_{1212}$ component of the  SP$07$ jet compared favourably against large-eddy simulation (LES) data of the same flow reported in Semiletov $\&$ Karabasov (2016); see Fig.(\ref{fig4_SPLpreds_comp}c).

Since the solution to non-parallel flow Green's function equation (\ref{Hyp3}) does not reduce uniformly to the parallel flow for the lower speed SP$03$ ($Ma=0.5$) jet, we found that a small anti-correlation region was required in turbulence model of ${R}_{1212}$ to achieve a reasonable estimation of the peak sound at these jet speeds (see Figs. \ref{fig4_SPLpreds} (d-f) $\&$ Fig. \ref{fig4_SPLpreds_comp}c).
An approximate composite model (Fig. \ref{fig4_SPLpreds_comp} a $\&$ b) that captures the effect of both parallel and non-parallel flow solutions to (\ref{Hyp3}) 
gave excellent agreement across a Strouhal number range %
but this obviously increases the empiricism in the application of the jet noise model, (\ref{I_low}). 
Future work will aim to compare the asymptotic theory for the propagator tensor that we have used in this paper to that determined by the full numerical solution of the ALEE (\ref{eq:GAA}) at $O(1)$ frequencies and acoustic Mach numbers. 

\section{Acknowledgement}

Computational resources from HPC2, Mississippi State University, are appreciated. MZA would like to thank Strathclyde University for financial support from the Chancellor's Fellowship.

%\disclaimer{Insert disclaimer text here.}

%%%%%%%%%% Insert bibliography here %%%%%%%%%%%%%%


\begin{thebibliography}{9}

 
\bibitem{1} Goldstein, M. E. 2003. A generalized acoustic analogy. \emph{J. Fluid Mech.}, \textbf{488}, pp. 315--333.

\bibitem{2} Crow, S. C., Champagne, F. H., 1971. Orderly Structures of Jet Turbulence. \emph{J. Fluid Mech.}, \textbf{48}, pp. 547--591

\bibitem{3} Lele, S. K., Nichols, J.W. 2014. A second golden age of aeroacoustics? \emph{Phil. Trans. R. Soc. A} 372: 20130321. http://dx.doi.org/10.1098/rsta.2013.0321.

\bibitem{4} Suzuki, T.  2013. Coherent noise sources of a subsonic round jet investigated using hydrodynamic and acoustic phased-microphone arrays. \emph{J. Fluid Mech.}, \textbf{730}, pp. 659--698.

\bibitem{5} Jordan, P., Colonius, T. (2013). Wave Packets and Turbulent Jet Noise.\emph{Ann. Rev. Fluid Mech.}. \textbf{45}, pp. 173-195.

\bibitem{6} Lighthill, M.J. 1952. On Sound Generated Aerodynamically: I. General Theory, \emph{Proc. R. Soc. Lon., A}, \textbf{211}, pp. 564-587.

\bibitem{7} Lilley, G. M. 1972.  On the Noise from Jets.,\textquotedblright\ \emph{AGARD CP-131}, pp. 13.1--13.12.


\bibitem{8} Karabasov, S. A., Afsar, M. Z., Hynes, T. P., Dowling, A.P., McMullan, W. A., Pokora, C. D., Page, G. J., McGuirk, J. J.2010. Jet Noise: Acoustic Analogy Informed by Large Eddy Simulation. \emph{AIAA J.}, \textbf{48}, No. 7, pp. 1312--1325.

\bibitem{9} Lele, S. K., Mendez, S., Ryu, J., Nichols, J., Shoeybi, M., Moin , P. 2010. Sources of high-speed jet nosie: analysis of LES data and modeling \emph{Procedia Engineering}, \textbf{6}, pp. 84--93.

\bibitem{10} Goldstein, M. E. and Leib, S.J. 2008.  The Aero-acoustics of slowly diverging supersonic jets. \emph{J. Fluid Mech.}, \textbf{600}, pp. 291--337.


\bibitem{11}  Afsar, M. Z. 2010.  Asymptotic properties of the overall sound pressure level of sub-sonic air jets using isotropy as a paradigm. \emph{J. Fluid Mech.}, \textbf{664}, pp. 510-539.

\bibitem{12} Afsar M. Z., Goldstein, M. E.,  Fagan, A. M (2011), Enthalpy flux/Momentum flux Coupling in the
Acoustic Spectrum of Heated Jets. \emph{AIAA J.}, \textbf{49}, No. 11, pp. 2522-2531.

\bibitem{13} Goldstein, M. E., Sescu, A., Afsar, M.Z. 2012,  Effect of non-parallel mean flow on the Green's function for predicting the low-frequency sound from turbulent air jets. \emph{J. Fluid Mech.}, \textbf{695}, pp. 199-234.

\bibitem{14}  Karabasov, S. A., Bogey, C and Hynes, T. P. 2013.  An investigation of the mechanisms of sound generation in initially laminar subsonic jets using the Goldstein acoustic analogy. \emph{J. Fluid Mech.}, \textbf{714}, pp. 24-57.

\bibitem{15} Afsar, M. Z., Sescu, A., Leib, S. J. 2016. Predictive Capability of Low Frequency Jet Noise using an Asymptotic Theory for the Adjoint Vector Green`s Function in Non-parallel Flow. \emph{22nd AIAA/CEAS Aeroacoustics Conference}, AIAA 2016-2804.


\bibitem{16} Bridges, J. 2006. Effect of heat on space-time correlations in jets. AIAA 2006-2534.

\bibitem{17} Tanna, H. K. 1977. An Experimental Study of Jet Noise. Part I: Turbulent Mixing Noise.  \emph{J. of Sound and Vib.}, \textbf{50}, No. 3, pp. 405-428.


\bibitem{18} Leib, S.J., Goldstein, M.E. 2011. Hybrid Source Model for Predicting High-Speed Jet Noise.  \emph{AIAA Journal}, \textbf{49}, No. 7. pp. 1324--1335.

\bibitem{19} Nelson, C. C. and Power, G.D. 2001. CHSSI Project CFD-7:The NPARC Alliance Flow Simulation System. \emph{AIAA Paper}, 2001-0594.

\bibitem{20} Nelson, C. C. 2010. An Overview of the NPARC Alliance's Wind-US Flow Solver. \emph{AIAA Paper}, 2010-27.

\bibitem{21}  Morse, P. M., Feshbach, H. 1953, Methods of Theoretical Physics. McGraw-Hill, USA.


\bibitem{22} Afsar, M. Z., Sescu, A., Sassanis, V.G. 2019. Effect of non-parallel mean flow on the acoustic spectrum of heated supersonic jets: explanation of `jet quietening'. \emph{Submitted to Phys. Fluids}.

\bibitem{23} Panchapakesan, N. R. and  Lumley,J. L. 1993.  Turbulence measurements in axisymmetric jets of air and helium. Part 1. Air jet. \emph{J. Fluid Mech}. \textbf{246}, pp. 197--223.

\bibitem{24}  Pope, S. B. 2000.  \textit{Turbulence}. Cambridge University Press, UK.

\bibitem{25}  Garebedian, P. R. 1998, \textit{Partial Differential Equations}. AMS Chelsea Publishing, Providence, Rhode Island, USA.

\bibitem{26} Van Dyke, M. 1975. \textit{Perturbation Methods in Fluid Mechanics}. The Parabolic Press, Stanford, California, USA. 

\bibitem{27} Pokora, C. D., McGuirk, J. J. 2015, Stereo-PIV measurements of spatio-temporal turbulence correlations in an axisymmetric jet. \emph{J. Fluid Mech.}, \textbf{778}, pp. 216--252.


\bibitem{28} Morris, P. and Zaman, K. 2010. Velocity Measurements in Jets with Application to Noise Source Modeling. \emph{J. Sound and Vib}., \textbf{329}, pp. 394-414.

\bibitem{29} Semiletov, V. A. and Karabasov, S. A. 2016. On the properties of fluctuating turbulent stress sources for high-speed jet noise. \emph{22nd AIAA/CEAS Aeroacoustics Conference, Aeroacoustics Conference}, AIAA 2016-2867. 

\bibitem{30} Harper-Bourne, M. 2003. Jet noise turbulence measurements. \emph{9th AIAA/CEAS Aero-acoustics conference}. AIAA 2003-3214.

\bibitem{31} Goldstein, M. E. and Leib, S. J. 2018. Azimuthal Source Noncompactness and Mode Coupling in Sound Radiation from High-Speed Axisymmetric Jets. \emph{AIAAJ}. {\bf 56}, pp. 3915-3926. 



\end{thebibliography}
\end{document}